\documentclass[journal,onecolumn]{IEEEtran}

\usepackage{amsmath,amssymb,amsfonts}
\usepackage[T1]{fontenc}
\usepackage[utf8]{inputenc}
\usepackage{graphicx}
\usepackage[export]{adjustbox}
\usepackage{booktabs}
\usepackage{multirow}
\usepackage{array}
\usepackage{float}
\usepackage{cite}
\usepackage{url}
\usepackage[unicode]{hyperref}
\graphicspath{{mathpix_assets/}}
\DeclareUnicodeCharacter{2713}{\checkmark}
\DeclareUnicodeCharacter{00D7}{\ensuremath{\times}}

\title{A Flow Model for the Electrified Railway-Power Grid Hybrid Asymmetric Coupled System and its Linearized Method}
\author{Qiao Zhang, Zhigang Liu,~\IEEEmembership{Fellow,~IEEE}, Shibin Gao, Zhenzu Liu, Xiangyu Meng, Yunchuan Deng, Guinan Zhang, Bing Lu, and Yanming Lu%
\thanks{This work was supported by the Science and Technology Project of Tibet Autonomous Region (XZ202402ZD0003). Corresponding author: Xiangyu Meng.}%
\thanks{Qiao Zhang, Zhigang Liu, Shibin Gao, Zhenzu Liu, Xiangyu Meng, Bing Lu, and Yanming Lu are with the School of Electrical Engineering, Southwest Jiaotong University, Chengdu 611756, China (e-mail: zhangqiao\_jq@163.com; liuzg\_cd@126.com; gao\_shi\_bin@126.com; lzzswjtu@163.com; mengxy55@126.com; lbj\_swjtu@163.com; yanmingll@foxmail.com).}%
\thanks{Yunchuan Deng is with China Railway Eryuan Engineering Group Co., Ltd., Chengdu, China (e-mail: dengdeng\_10@126.com).}%
\thanks{Guinan Zhang is with China Academy of Railway Sciences Group Co., Ltd., Beijing, China (e-mail: zgn\_2008@126.com).}}

\begin{document}
\maketitle

\begin{abstract}
In mountainous regions where traction loads constitute a significant portion of a long-chain weak power grid (PG) with sustainable energy, the interaction between the traction power supply system and the PG becomes increasingly evident. The integrated power flow calculation (PFC) method and its linearized model are quite important for the PG - traction network (TN) joint planning. However, existing research on the port load characteristics of the EMUs and the connection angle characteristics of traction transformers is insufficient, and there is a lack of effective methods for PFC or linearized PFC in systems that couple the PG with the traction network. To fill this gap, this paper proposes an integrated PFC model for the AT TN - PG coupled system, along with a linearized method. Firstly, according to the relationship of the phases between the PG and the AT traction network, the node admittance matrix of the coupled system has been constructed. Then, the issue of power injection equations being unable to deal with the EMUs port load is resolved by merging the contact line node and the rail node. Subsequently, the integrated PFC equations for the coupling system are established. Next, a hybrid phase linear decoupled power flow model for the coupling system is developed, employing the correspondence between the phases of the PG and the TN, as well as the phase angle differences among various nodes and branches. Numerical simulations conducted in a specific region demonstrate the necessity of an integrated PFC for the coupled system and validate both the accuracy and efficiency of the linearized model.
\end{abstract}


\section*{Nomenclature}
\begin{center}
\centering
\scriptsize
\setlength{\tabcolsep}{3pt}
\renewcommand{\arraystretch}{1.08}
\begin{tabular}{@{}>{\raggedright\arraybackslash}p{0.075\textwidth}>{\raggedright\arraybackslash}p{0.37\textwidth}>{\raggedright\arraybackslash}p{0.075\textwidth}>{\raggedright\arraybackslash}p{0.37\textwidth}@{}}
AT & autotransformer & PG & power grid \\
TN & traction network & PFC & power flow calculation \\
TSS & traction power supply system & EMUs & electric multiple units \\
TNPGS & traction network--power grid coupled system & ATPGS & AT traction network--power grid coupled system \\
T & transmission phase of the traction network & F & negative feeder phase of the traction network \\
R & return phase of the traction network & PCC & point of common coupling \\
HPLD & hybrid phase linear decoupled & NR & Newton--Raphson \\
$V_m$ & voltage magnitude & $V_a$ & voltage angle \\
DUM & average error of the voltage magnitude & DUA & average error of the voltage angle \\
$U_{km}^{\mathrm{HPLD}}$ & voltage magnitude of the $k$th node of phase $m$ calculated by the HPLD method & $U_{km}^{\mathrm{NR}}$ & voltage magnitude of the $k$th node of phase $m$ calculated by the NR method \\
$\theta_{km}^{\mathrm{HPLD}}$ & voltage angle of the $k$th node of phase $m$ calculated by the HPLD method & $\theta_{km}^{\mathrm{NR}}$ & voltage angle of the $k$th node of phase $m$ calculated by the NR method \\
ABPG & phase A branch on the power grid side & BBPG & phase B branch on the power grid side \\
CBPG & phase C branch on the power grid side & TBTT & phase T branch of the traction transformer \\
AD & angle difference & AAD & average value of angle difference \\
MAD & maximum value of angle difference & AAE & average value of angle error \\
MAE & maximum value of angle error & VAD & voltage angle difference \\
\end{tabular}
\end{center}

\section*{I. Introduction}
WITH the development of economic levels, the planning and construction of railways in high-altitude mountainous areas have become increasingly urgent. However, these regions often experience low power loads and typically exhibit a long-chain weak power grid structure, with a small short-circuit capacity at the end of the chain. Once the railway is completed, the proportion of the traction load may reach $20-30 \%$. At this point, the coupling between the traction network and the power grid becomes more stringent, as shown in Fig. 1. Traditional planning of TSSs treats the upper-level power grid as a slack node, which is difficult to apply in the long-chain weak power grid structures of mountainous areas. This is because the nodes at the front end of the chain must not only bear the traction load of their own connections but also withstand the impact of the loads from the rear end of the chain.

Moreover, railways in high-altitude mountainous areas often feature long slopes [1], which in some cases require the TSSs to adopt a bilateral power supply structure (as shown in Fig. 1), eliminating the electrical phase separation between traction arms to ensure the safe and stable operation of EMUs on long slopes [2]-[4]. In such cases, the TN and the PG form a circular coupling structure, making it challenging to conduct power flow analysis separately for the TN and the PG. Therefore, the planning of TSSs under the long-chain power grid structure in remote mountainous areas urgently requires an integrated flow calculation model for the TNPGS, as well as its linear flow model as an optimization tool. To this end, a systematic literature review will be conducted on the PFC models of power systems and TSSs, as well as their linearization methods.

\section*{A. Literature Survey}
The nonlinear PFC methods have reached a mature level in power system analysis, applicable to both single-phase and three-phase PFC. This includes traditional algorithms such as the Newton-Raphson method, PQ decomposition method, and Gauss-Seidel method. These algorithms are not only suitable for PFC in large power systems but are also widely used in distribution networks. Most existing algorithms are based on known node injection power, using Vm and Va as iterative variables. Beyond classical algorithms such as the Newton-Raphson method, modern continuous optimization algorithms represented by interior-point methods (IPM) and trust-region methods (TRM) have become important tools for handling large-scale, tightly-constrained, and ill-conditioned systems.

The interior-point method handles inequality constraints by introducing barrier functions and searches for the optimum within the interior of the feasible region. Due to its excellent convergence and polynomial-time complexity, it has become one of the mainstream algorithms for solving optimal power flow (OPF) problems [5]-[6]. Its applications have expanded from basic models to multi-objective optimization and parallel accelerated computing [7]. The trust-region method, on the other hand, constructs a local approximate model and solves it within a dynamically adjusted trust region. It is renowned for its strong global convergence and robustness [8], making it particularly suitable for scenarios where traditional methods struggle to converge, such as power flow calculations in islanded microgrids without slack buses [9] or severely ill-conditioned systems [8]. Furthermore, hybrid algorithms like the trust-region interior-point method, which combine the advantages of both, have been proposed to simultaneously ensure robustness and computational efficiency [10].

However, in the TSSs, the EMUs are distributed between the contact line and the rails. The current is drawn from the contact line and then returned through the rails, which constitutes a port load. Therefore, it is impossible to specify the injection power at the nodes, making it challenging to directly apply the aforementioned traditional algorithms to TSSs.

To address this specific issue, a sequence linear method [11] and its derived algorithms [12]-[14] based on current have been specially developed to meet the needs of PFC in TSSs. In\\[0pt]
addition, the node analysis method [15] and the current-based Newton-Raphson methods [16][17] have also been proposed for PFC in TSSs. However, these methods have limitations when dealing with PV nodes and are typically only effective for handling the slack node, which makes them unsuitable for PFC in power systems. Therefore, there is an urgent need for a PFC model that can comprehensively consider the unique port loads of EMUs in the TNPGS to fill this gap.

In addition, the aforementioned PFC models for TSSs are all nonlinear models, with few linear models reported. In contrast, linear flow calculation models in power systems have seen considerable development and can be mainly divided into single-phase and three-phase linear models.

The single-phase linear model is most well-known through the classic DC power flow method [18], but it has become difficult to meet the demands of modern power system analysis due to its inability to consider voltage magnitudes and reactive power. To solve this challenge, an improvement to the DC method was proposed in [19], which takes into account both active and reactive power by establishing a linear power flow model using voltage magnitude squared and the product of voltage magnitude squared and phase angle as variables. However, coupling between Vm and Va still exists. Thus, a single-phase linear algorithm suitable for radial networks/distribution systems was introduced in [20] but did not consider PV nodes. Progress has been made in [21]-[24] towards linearizing the state equation, resulting in a linear solution model for Vm and Va. Nevertheless, the linear equations for node voltages do not consider phase shifts, which limits their applicability in distribution networks or TSSs. [25]constructed a refined and comprehensive single-phase approximate linear model, considering transformer tap ratios and phase shifts through logarithmic transformation and first-order Taylor expansion. In conclusion, these single-phase linear models not only linearize the power flow equations but also take into account important factors such as voltage magnitudes and reactive power, thus compensating for the limitations of traditional DC models. However, these models have certain limitations when directly applied to distribution networks or TSSs with different tap-changing transformers.

The existing three-phase linearized power flow models are mainly focused on distribution networks. [26] based on the DistFlow equations, constructed a three-phase linearized power flow model suitable for radial networks utilizing voltage squared and current magnitude as variables, but its application on loop networks has limitations. [27] proposed a generalized LinDistFlow power flow analysis model, yet lacking detailed analysis of voltage angles. In response, [28] exploited the characteristic of 120 -degree phase difference in three-phase systems to devise a linearized method suitable for three-phase distribution systems. Furthermore, [29] employed ensemble learning to develop a three-phase power flow linear approximation method. Additionally, [30] established a linear relationship between branch power and logarithm of voltage through logarithmic transformation and first-order Taylor expansion. While [28]-[30] considered PV nodes, Y-shaped loads, and delta-shaped loads, they did not account for different\\
transformer connection angle variations, such as $30^{\circ}$ or $60^{\circ}$. On the other hand, [31] introduced a distribution network linearized power flow model based on curve fitting, while [32] utilized the Z-bus iterative algorithm to derive a linearized distribution network power flow model through first-order Taylor expansion. These two models can accommodate both Y-shaped loads and delta-shaped loads, regardless of transformer connection configurations. Nonetheless, these models cannot handle PV nodes, thus they do not apply to TNs.

Furthermore, research on railway-power grid integration has deepened around coupling mechanisms, modeling methods, optimal control, and new energy synergy. In terms of coupling impact mechanisms, existing studies have analyzed traction load-induced negative sequences, harmonics, and other power quality issues [33], quantified their spatiotemporal impact on grid voltage [34], and addressed waveform distortion propagation in low-frequency electrification scenarios [35]. Modeling technologies have advanced with ADPSS-based hybrid simulation [36] and the Compressed Newton-Raphson method [37] improving accuracy and efficiency, yet existing linear models fail to consider the $30^{\circ}$ phase shift of V/x6 traction transformers and PV nodes in ATPGS. Optimal control has formed diverse paths, including centralized compensation for weak grids [38], energy storage-integrated conditioners [33], and new energy integration schemes [39]. Grid support capacity evaluation has also emerged as a hotspot [40].

In conclusion, existing distribution network linear models [28]-[32] handle unbalance but cannot simultaneously accommodate transformer connection angles and PV nodes, while railway-side studies focus on independent traction network modeling [16], ignoring grid coupling nonlinearities.

\section*{B. Contribution}
This paper primarily addresses a fundamental problem at the modeling level: namely, how to establish an accurate integrated nonlinear power flow model for the ATPGS (Section II). This model correctly characterizes the system's asymmetric coupling characteristics and EMU port loads, forming the physical foundation for all subsequent analyses-whether solved using the NR method or more advanced algorithms. Building upon this foundation, and to tackle the bottleneck of "massive-scenario computational efficiency" encountered when applying this model in planning, this paper further proposes a planning-oriented customized linear model, HPLD. The main contributions are threefold.\\
(1) In view of the power flow node injection power equation cannot deal with EMUs port load, this paper proposes a method of merging contact line node and rail node to solve this problem. On this basis, fully utilizing the correlation between the phases of the power grid and the traction network, an integrated PFC model for the ATPGS is established.\\
(2) To fill the gap of existing linear models that are not applicable to the ATPGS, the hybrid phase linear decoupled (HPLD) power flow method for ATPGS is derived, considering different traction transformer tap angles by assuming constant voltage angle differences between different phases at the same node and across different branches. This linear model

\begin{figure}[H]
\begin{center}
  \includegraphics[alt={},max width=\columnwidth]{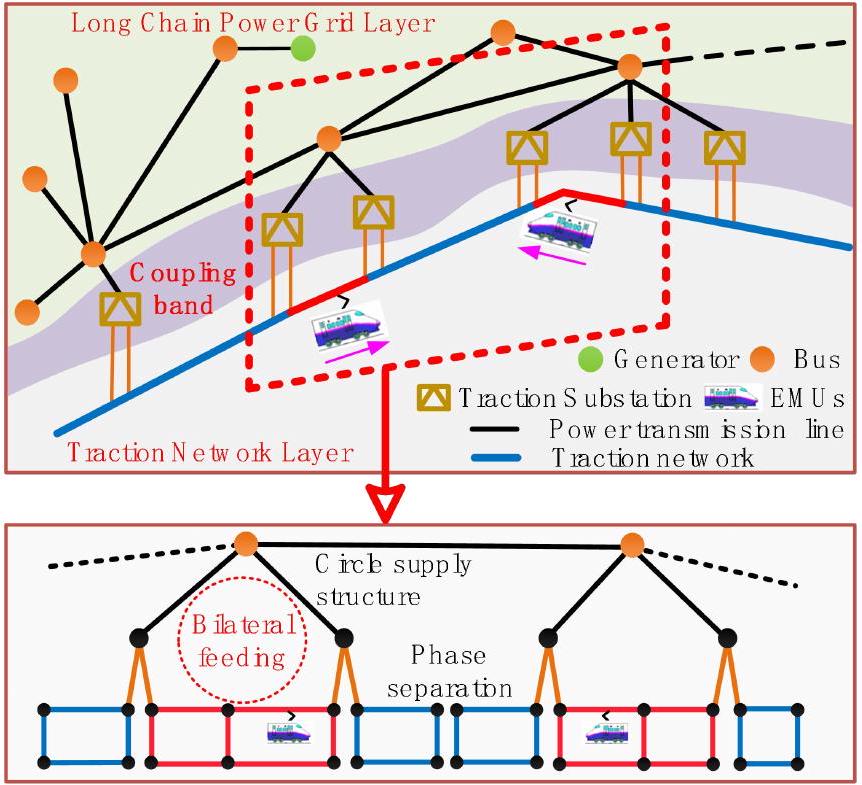}
\caption{Diagram of long chain and bilateral circle supply structure in TNPGS.}
\end{center}
\end{figure}

\begin{figure}[H]
\begin{center}
  \includegraphics[alt={},max width=\columnwidth,max height=0.72\textheight,keepaspectratio]{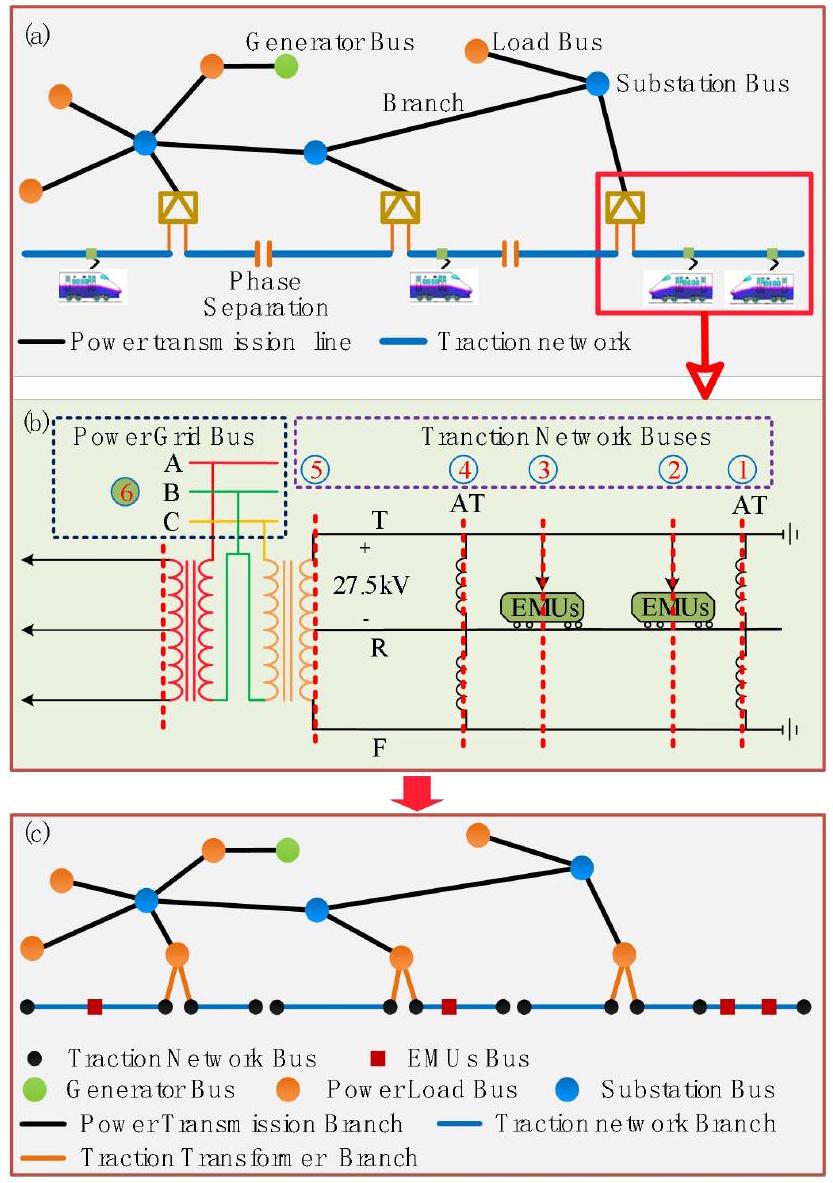}
\caption{Schematic diagram of topological modeling process for the ATPGS.}
\end{center}
\end{figure}

simultaneously incorporates the voltage magnitude, the voltage angle, transformer tap angles, and PV nodes.\\
(3) The proposed integrated PFC method of the ATPGS and its linearized model HPLD can comprehensively analyze the interaction between the TSSs of the whole railway and the power system, such as the influence of different EMUs

\begin{figure}[H]
\begin{center}
  \includegraphics[alt={},max width=\columnwidth]{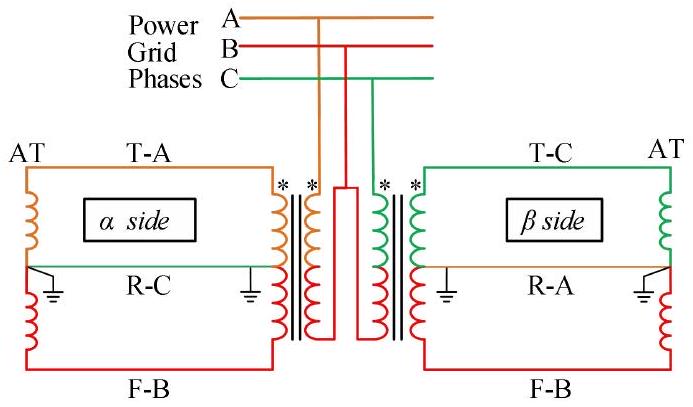}
\caption{Phase sequence correspondence between the PG side and TN side of the V/x6 traction transformer.}
\end{center}
\end{figure}

\begin{figure}[H]
\begin{center}
  \includegraphics[alt={},max width=\columnwidth]{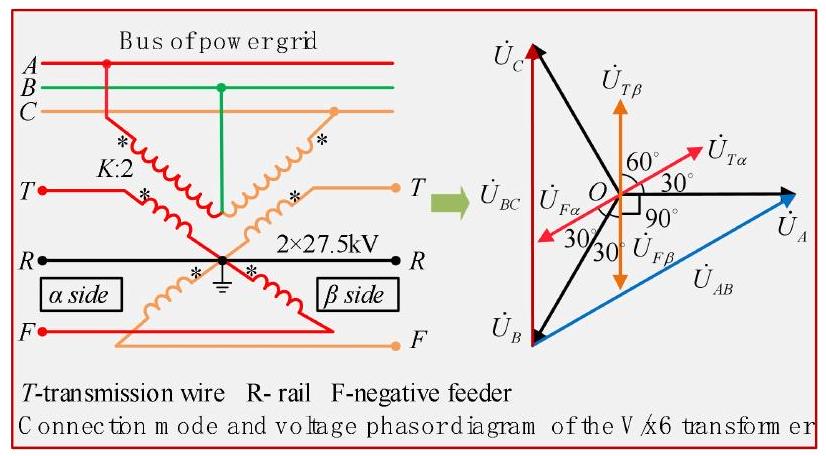}
\caption{Schematic diagram of the wire connection and the voltage phasor of the V/x6 traction transformer.}
\end{center}
\end{figure}

\begin{table}[H]
\begin{center}
\caption{Voltage Angle of PG Side and TN Side}
\begin{tabular}{ccccccc}
\hline
\multicolumn{3}{c}{PG side} & \multicolumn{4}{c}{TN side} \\
\hline
Phase & Angle $\left({ }^{\circ}\right)$ & Phase & $\mathrm{V} / \mathrm{x} 0$ & $\mathrm{~V} / \mathrm{x} 6$ & Scott/x &  \\
$\boldsymbol{U A}$ & 0 & $\boldsymbol{U T} \boldsymbol{\alpha}$ & -150 & 30 & 30 &  \\
$\boldsymbol{U B}$ & -120 & $\boldsymbol{U F} \boldsymbol{\alpha}$ & 30 & -150 & -150 &  \\
$\boldsymbol{U C}$ & 120 & $\boldsymbol{U T} \boldsymbol{\beta}$ & -90 & 90 & 120 &  \\
- & - & $\boldsymbol{U F} \boldsymbol{\beta}$ & 90 & -90 & -60 &  \\
\hline
\end{tabular}
\end{center}
\end{table}

operating conditions on the voltage unbalance of the long chain weak network. In addition, in the process of solving the programming model, the linear model can easily find the optimal solution. Thus, they can provide powerful analysis tools for the planning of ATPGS in remote mountain area.

\section*{C. Paper Organization}
The remainder of the paper is organized as follows. Construction of the topological structure and integrated power flow equation for ATPGS are described in Section II. The HPLD method is described in Section III. Section IV shows the handling of PV and slack nodes in three-phase asymmetric power flow calculations. The Case analysis is shown in Section V. Finally, conclusions are drawn in Section VI.

\section*{II. Construction of Topological Structure and Power Flow Equation for ATPGS}
\section*{A. Topological Structure Modeling of ATPGS}
In a regional mountain PG with a high proportion of traction loads, the PG and the TN naturally form a coupled system, as shown in Fig. 1. It is easy to distinguish between the PG and the TN layer, which are coupled through traction transformers.

The topology of the power system is naturally formed, with transmission lines and transformers acting as branches. Power\\
plants, high and low side of substations, and loads serve as nodes. However, in TSSs, due to the continuous forward movement of EMUs, the topology exhibits dynamic time-varying characteristics, as shown in Fig. 2(a). For such systems, a cutting method is used as a feasible approach to establish the topology structure. In the AT TSSs, the EMUs and AT stations are each regarded as nodes, as depicted in Fig. 2(b). Two EMUs operate on the right supply arm simultaneously, incorporating two AT stations, thereby forming a chain structure from node 1 to node 4. A schematic of the chain structure model is illustrated in Fig. 1, where the EMUs are considered as power sources. Furthermore, the high-voltage and low-voltage busbars of the traction transformers are also regarded as nodes, as shown by nodes 5 and 6 in Fig. 2(a). Through the transformation indicated in Fig. 2(b), the coupling system shown in Fig. 2(a) can be converted into the pure topology structure depicted in Fig. 2(c), which includes generator nodes, power load nodes, substation nodes, traction network nodes, and EMUs nodes. The generator node is treated as a PV or slack node, while the others are treated as PQ nodes. As the EMUs move, the topology changes with their positions, thereby forming a dynamic topology structure.

\section*{B. Phase Sequence Correspondence between PG and AT TN}
The power system is a three-phase transmission system. The AT TSS is essentially a two-phase transmission system, where T and F serve the role of transmission, while the rails serve the purpose of return flow, named the return phase (R). Therefore, it is also represented in the form of "three-phase transmission." Additionally, the voltage angle of the transmission wire on the left and right power supply arms of the TN are generally different and depend on the type of traction transformer. As a result, the three-phase power flow in this coupling system is often unbalanced. From the perspective of the entire system, when studying the coupling characteristics of power flows on both the PG and the TN sides, it is necessary to consider the coupling system flow equations that include both sides. Thus, a correspondence must be established between the phase sequences $\mathrm{A}, \mathrm{B}$, and C on the PG side and the phase sequences $\mathrm{T}, \mathrm{F}$, and R on the TN side. The principle of phase sequence correspondence should be based on the matching ends of the transformer windings. Fig. 3 illustrates the phase sequence correspondence of the V/x6 traction transformer. Different transformers lead to phase differences in different supplying arms, as detailed in Table I. The schematic diagram of the wire connection and the voltage phasor of the V/x6 traction transformer are shown in Fig. 4. It can be observed that the voltage phases on the $\alpha$ side and $\beta$ side are significantly different - the voltage angle of T on the $\alpha$ side leads the phase A by $30^{\circ}$, while it is $90^{\circ}$ on the $\beta$ side. Furthermore, the phase difference between the T and the F on both $\alpha$ and $\beta$ sides is $180^{\circ}$. Furthermore, it should be noted that in order to reduce the overall system unbalance, the phase sequence of the traction transformer when connected to the PG is not always in the ABC order. This will also result in different Va on the low-voltage side supply arms. For specific details, please refer to the Table S-I of supplementary materials. These characteristics differ

\begin{figure}[H]
\begin{center}
  \includegraphics[alt={},max width=\columnwidth]{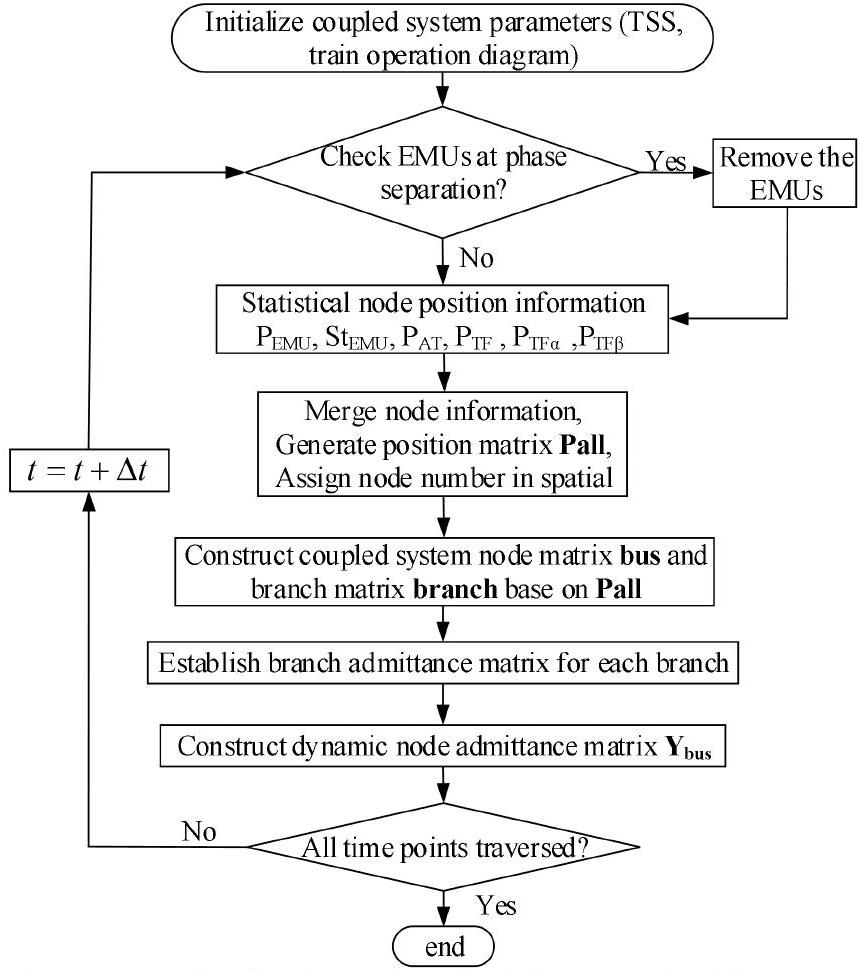}
\caption{Construction flowchart of the node admittance matrix of the ATPGS}
\end{center}
\end{figure}

\begin{figure}[H]
\begin{center}
  \includegraphics[alt={},max width=\columnwidth]{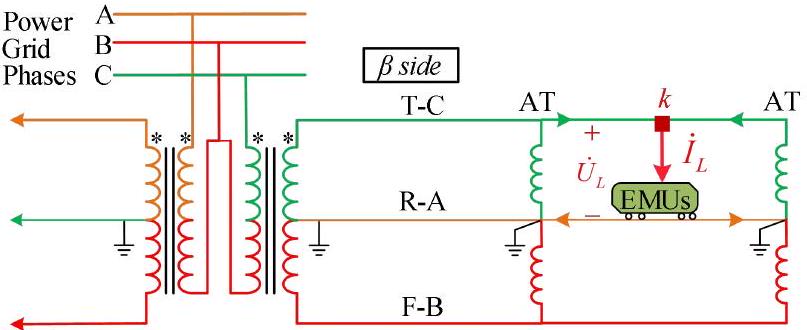}
\caption{Simplified ATPGS power supply structure diagram.}
\end{center}
\end{figure}

from those of distribution networks and will provide key information for the initial values setting for PFC and linearizing the power flow equations of the ATPGS.

\section*{C. Node Admittance Matrix Modeling of ATPGS}
The modeling of the nodal admittance matrix for the ATPGS is similar to that of a three-phase power system. The overall approach involves deriving the branch admittance matrices for the various components and then constructing the nodal admittance matrix for the entire topological network based on the connection relationships between the nodes. Therefore, in this coupled system, the components for which branch admittance matrices need to be established include the power transmission lines, transformers on the PG side, TN [15][41], traction transformer [41] ((S1) of the supplementary material), and AT [41]. The positions of nodes on the power grid side are fixed, while the positions of EMUs are constantly changing, leading to continuous variations in the node admittance matrix of the entire system. And each traction power supply system unit consists of two power supply arms. The specific node admittance matrix construction flowchart is shown in Fig. 5 and detailed steps are as follows:

Step1: Screen Valid EMUs. Based on the train operation diagram, determine whether there are EMUs in the phase\\
separation zone at time $t$. If yes, exclude these EMUs; otherwise, proceed to the next step.

Step2: Collect Node Position Information. Count the position information of each node at time $t$ using the train operation diagram. Specifically, for each traction power supply system unit: 1) The position information of EMUs is denoted by $\mathrm{P}_{\mathrm{EMU}}$ (e.g., $\mathrm{P}_{\mathrm{EMU}}=[5,15]$ ), where the values represent kilometer markers). 2) The uplink/downlink status of EMUs is indicated by $\mathrm{St}_{\text{EMU }}$, with elements being 1 (uplink) or -1 (downlink). 3) The position information of Auto Transformer (AT) stations or parallel points is represented by $\mathrm{P}_{\mathrm{AT}}$ (e.g., $\mathrm{P}_{\mathrm{AT}}=$ [0,10,20]). 4) The position information of traction substations is denoted by $\mathrm{P}_{\mathrm{TF}}$, and the position information of the exit points of the left and right power supply arms is represented by $\mathrm{P}_{\mathrm{TF} \alpha}$ and $\mathrm{P}_{\mathrm{TF} \beta}$, respectively.

Step3: Generate Merged Position Matrix. Merge the node information collected in Step 2 to form the position matrix $\mathrm{P}_{\text{All }}$, and assign node numbers to EMUs, AT stations, parallel points, and traction substations in spatial order.

Step4: Construct Coupled System Node Matrix. Based on the node numbers in Step 3, combined with generator nodes and other nodes on the power grid side, construct the node matrix bus of the coupled system. Each element in the $\mathrm{P}_{\mathrm{All}}$ matrix constitutes a node, which is set as a load node.

Step5: Establish Branch Connection Matrix. Adjacent nodes on the power grid side form a branch, and adjacent nodes in the $\mathrm{P}_{\text{All }}$ matrix also form a branch, thereby constructing the branch connection matrix branch.

Step6: Build Branch Admittance Matrix. Construct the branch admittance matrix for each branch in the system.

Step7: Construct Dynamic Node Admittance Matrix. Based on the branch connection relationships, the EMU uplink/downlink matrix $\mathrm{St}_{\text{emu }}$, and the branch admittance matrix, construct the dynamic node admittance matrix $\mathrm{Y}_{\text{bus }}$ of the coupled system.

Step8: Time Iteration Check. Determine whether all time moments have been traversed. If yes, terminate the process; otherwise, update the time to $t=t+\Delta t$ and return to Step 1.

It should be noted that since the $\mathrm{P}_{\text{All }}$ matrix changes with time, the node admittance matrix $\mathrm{Y}_{\text{bus }}$ also varies over time. Therefore, $\mathrm{Y}_{\text{bus }}$ is the dynamic node admittance matrix of the coupled system.

\section*{D. Power Flow Equations Construction of ATPGS}
When considering the coupled characteristics of the ATPGS, it is necessary to take into account the load flow equations consisting of the power grid and the traction network. The simplified ATPGS is shown in Fig. 6. Taking the active power injection equation of phase A as an example, the equation consists of two parts, as shown in (1).

\begin{equation*}
P_{i a}=U_{i a}\binom{\sum_{j=1}^{N_{P G}} \sum_{m \in \Omega_{P G}}\left(G_{i j a m} \cos \theta_{i a j m}+B_{i j a m} \sin \theta_{i a j m}\right) U_{j m}+}{\sum_{j=N_{P G}+1}^{n} \sum_{m \in \Omega_{T N}}\left(G_{i j a m} \cos \theta_{i a j m}+B_{i j a m} \sin \theta_{i a j m}\right) U_{j m}} \tag{1}
\end{equation*}

where $n$ represents the number of all nodes. $\boldsymbol{Y}=\boldsymbol{G}+j \boldsymbol{B}$ is the node admittance matrix. $N_{P G}$ and $\Omega_{P G}$ represent the number of nodes and the set of phases belonging to PG, $\Omega_{P G}=\{\mathrm{A}, \mathrm{B}, \mathrm{C}\}$. $\Omega_{T N}$ represents the set of phases belonging to TSS, $\Omega_{T N}=\{\mathrm{T}, \mathrm{R}$,\\
$\mathrm{F}\}$. It should be specifically noted that the R serves only a reflux function, rather than a transmission phase. However, due to its equal length with T and F and the existence of mutual inductance, it is necessary to consider it in the equation.

As indicated by (1), it is known that the number of transmitted phases on the PG side and the TSS side are the same, thus allowing them to be merged into the form shown in (2). However, attention needs to be paid to the correspondence between the phase sequence on the PG side and the TSS side, as exemplified by the relationship T-A, F-B, and R-C on the $\alpha$ side of the V/x6 traction transformer.

\begin{equation*}
P_{i a}=U_{i a} \sum_{j=1}^{n} \sum_{m \in\{a, b, c\}}\left(G_{i j a m} \cos \theta_{i a j m}+B_{i j a m} \sin \theta_{i a j m}\right) U_{j m} \tag{2}
\end{equation*}

Through analyzing (2), it is revealed that the ATPGS can be regarded as a strongly coupled nonlinear system of hybrid phases. Strong coupling relationships exist among the Vm, Va, and their interdependencies across different phases.

It should be particularly noted that in the traction substation and the AT substation, both the neutral points of the traction transformer and the secondary side of the AT need to be grounded. Assuming the set of these nodes is $\Omega_{0}=\left\{g_{1}, g_{2}, \ldots\right.$, $\left.g_{k}\right\}$. Therefore, during the PFC, the voltages at these points are zero and do not participate in the iterative solution process. Assuming their phase sequence is X , we have

\begin{equation*}
U_{g_{k} X}=0, g_{k} \in \Omega_{0} \tag{3}
\end{equation*}

where, X can be any phase among $\mathrm{A}, \mathrm{B}$, and C , determined by the connection sequence on the high-voltage side of the traction transformer. $g_{k}$ is the number of the grounded node.

\section*{E. Power Injection Processing for EMUs Node}
The EMUs obtain electrical energy from the contact wire via the pantograph and flow it back through the rails, as illustrated in Fig. 6. Therefore, the trains experience a voltage between the contact wire and the rails. Assuming the numbered node of the EMUs is $k$ and it is on the $\alpha$ side. The current flowing through the EMUs is $\dot{I}_{L}$ and the voltage of the EMUs is $\dot{U}_{L}$, then the complex power consumed by the EMUs is

\begin{equation*}
S_{L}=U_{L} I_{L}^{*} \tag{4}
\end{equation*}

Since the voltage of EMUs is not equal to the contact wire voltage or the rail voltage, the power of the EMUs is not equal to the injected power from the contact wire or the rail, that is,

\begin{align*}
& U_{L}=U_{k T}-U_{k R} \neq U_{k T} \neq U_{k R} \Rightarrow U_{L} I_{L}^{*} \neq U_{k T} I_{L}^{*} \neq U_{k R} I_{L}^{*}  \tag{5}\\
& \Rightarrow S_{L} \neq S_{k T} \neq S_{k R}
\end{align*}

where $\dot{U}_{k T}$ denotes the voltage of T in node $k . \dot{U}_{k R}$ denotes the voltage of R in node $k . \dot{S}_{k T}$ is the injection complex power of T in node $k . \dot{S}_{k R}$ is the injection complex power of T in node $k$.

Thus, the complex power of the EMUs cannot be directly substituted into the nodal power injection equation (2) without transformation. Here, a method of node merging is presented to address this issue. The detailed steps are as follows.

Step1: Construct the matrix equation based on Ohm's law for the ATPGS:

\[
\left\{\begin{array}{l}
\boldsymbol{U}=\boldsymbol{Z} \boldsymbol{I}, \boldsymbol{U}=\left[\begin{array}{llllll}
U_{1 a} & U_{1 b} & U_{1 c} & \cdots & U_{n a} & U_{n b} \\
\cdot & \cdot & U_{n c}
\end{array}\right]^{T}  \tag{6}\\
\boldsymbol{I}=\left[\begin{array}{llllll}
I_{1 a} & I_{1 b} & I_{1 c} & \cdots & I_{n a} & I_{n b}
\end{array} I_{n c}\right.
\end{array}\right]^{T} .
\]

where $\boldsymbol{U}$ is the node voltage vector, $\boldsymbol{I}$ is the node injection current vector, and $\boldsymbol{Z}$ is the impedance matrix, and its specific form is detailed in formulation (S1) of the supplementary materials.

Step2: List the voltage and current relationships of EMUs nodes $k$, as shown in (7) and (8).

\begin{align*}
& -I_{k a}=I_{k c}=I_{L}  \tag{7}\\
& U_{k a}-U_{k c}=U_{L} \tag{8}
\end{align*}

Step3: Rewrite $\dot{U}_{L}$ by Substituting (6) and (7) into (8), we have

\begin{align*}
U_{L} & =\left(Z_{k 1 a a}-Z_{k 1 a c}\right) I_{1 a}+\left(Z_{k 1 a b}-Z_{k 1 b c}\right) I_{1 b}+\left(Z_{k 1 a c}-Z_{k 1 c c}\right) I_{1 c}+\cdot \\
& +\left(-\left(Z_{k k a a}-Z_{k k a c}\right)+\left(Z_{k k a b}-Z_{k k b c}\right)\right) I_{L}+\left(Z_{k k a c}-Z_{k k c c}\right) I_{k b b}+\cdot  \tag{9}\\
& +\left(Z_{k n a a}-Z_{k n a c}\right) I_{n a}+\left(Z_{k n a b}-Z_{k n b c}\right) I_{n b}+\left(Z_{k n a c}-Z_{k n c c}\right) I_{n c}
\end{align*}

Step4: In (6), replacing $\dot{U}_{k a}$ and $\dot{U}_{k c}$ with $\dot{U}_{L}$ and replacing $\dot{I}_{k a}$ and $\dot{I}_{k c}$ with $\dot{I}_{L}$, a new Ohm's law equation (10) can be obtained with one fewer dimension than before.

\begin{align*}
\boldsymbol{U}^{\prime} & =\boldsymbol{Z} \boldsymbol{I}^{\prime} \\
\boldsymbol{U}^{\prime} & =\left[\begin{array}{llllllllll}
\dot{U}_{1 a} & \dot{U}_{1 b} & \dot{U}_{1 c} & \cdots & \dot{U}_{L} & \dot{U}_{k b} & \cdots & \dot{U}_{n a} & \dot{U}_{n b} & \dot{U}_{n c}
\end{array}\right]^{T}  \tag{10}\\
\boldsymbol{I}^{\prime} & =\left[\begin{array}{llllllllll}
\dot{I}_{1 a} & \dot{I}_{1 b} & \dot{I}_{1 c} & \cdots & \dot{I}_{L} & \dot{I}_{k b} & \cdots & \dot{I}_{n a} & \dot{I}_{n b} & \dot{I}_{n c}
\end{array}\right]^{T}
\end{align*}

where $\boldsymbol{U}^{\prime}$ donates the transformed node voltage vector. $\boldsymbol{I}^{\prime}$ donates the transformed node injection current vector. $\boldsymbol{Z}^{\prime}$ donates the transformed impedance matrix, and its specific form is detailed in (S2) of the supplementary materials.

Step5: A new nodal admittance matrix $\boldsymbol{Y}^{\prime}=\boldsymbol{G}^{\prime}+j \boldsymbol{B}^{\prime}$ can be obtained by the inverse of the impedance matrix $\boldsymbol{Z}^{\prime}$. In this way, the power of the EMUs can be directly substituted into (2) for the iteration calculation using the NR method. Therefore, the integrated PFC method considering the EMUs port load of the ATPGS has been established.

\section*{III. Hybrid Phase Linear Decoupled Power Flow Equations of ATPGS}
\section*{A. Motivation of HPLD}
A direct first-order Taylor expansion on the original coupled power flow equations (2) was initially considered. However, this approach encounters significant theoretical obstacles. The presence of special traction transformers (e.g., $\mathrm{V} / \mathrm{x}$ ) and grounded nodes (e.g., neutral points) introduces numerous conditional branches in the derivative calculations, leading to prohibitively complex and piecewise Jacobian expressions. This complexity renders a direct Taylor expansion impractical for an efficient, general-purpose linear model. Therefore, a feasible two-stage linearization method (HPLD) is proposed as follows.

\section*{B. Stage 1: the Phase Assumption of the Same Node}
In a symmetrical system, the VAD within the same node remains constant. For instance, in a three-phase power system where the voltage angles of each phase differ by $120^{\circ}$, while the VAD between T and F of the AT TN powered by a V/x6 transformer is $180^{\circ}$. Therefore, to decouple the Va of different phases at the same node, the VAD between different phases at the same node can be assumed to be a constant, namely

Assumption 1:

\begin{equation*}
\gamma_{i a i m} \approx \theta_{i a}-\theta_{i m}, m \in \Omega=\{a, b, c\} \tag{11}
\end{equation*}

where $y_{\text{iaim }}$ is a constant angle, which can be determined based on the phasor diagram, as shown in Fig. 4.

Thus, the active power injection equation with angle decoupled can be obtained by substituting (11) to (2), that is

\begin{align*}
& P_{i a} \approx U_{i a} \sum_{j=1}^{n} \sum_{m \in \Omega}\binom{G_{i j a m}^{\prime} \cos \left(\theta_{i m}-\theta_{j m}+\gamma_{i a i m}\right)}{+B_{i j a m}^{\prime} \sin \left(\theta_{i m}-\theta_{j m}+\gamma_{i a i m}\right)} U_{j m} \\
& =U_{i a} \sum_{j=1}^{n} \sum_{m \in \Omega}\binom{\left(\cos \gamma_{i a i m} G_{i j a m}^{\prime}+\sin \gamma_{i a i m} B_{i j a m}^{\prime}\right) \cos \theta_{i m j m}}{+\left(-\sin \gamma_{i a i m} G_{i j a m}^{\prime}+\cos \gamma_{i a i m} B_{i j a m}^{\prime}\right) \sin \theta_{i m j m}} U_{j m}  \tag{12}\\
& =U_{i a} \sum_{j=1}^{n} \sum_{m \in \Omega}\left(H_{i j a m} \cos \theta_{i m j m}+F_{i j a m} \sin \theta_{i m j m}\right) U_{j m}
\end{align*}

where

\begin{gather*}
H_{i j a m}=\left\{\begin{array}{c}
-h_{i j a m} j \neq i \\
h_{i i a m}+\sum_{k=1, k \neq i}^{n} h_{i k a m} j=i
\end{array}, F_{i j a m}=\left\{\begin{array}{c}
-f_{i j a m} j \neq i \\
f_{i i a m}+\sum_{k=1, k \neq i}^{n} f_{i k a m} j=i
\end{array}\right.\right.  \tag{13}\\
\left\{\begin{array}{c}
h_{i j a m}=\left(\cos \gamma_{i a i m} g_{i j a m}^{\prime}+\sin \gamma_{i a i m} b_{i j a m}^{\prime}\right) \\
f_{i j a m}=\left(\cos \gamma_{i a i m} b_{i j a m}^{\prime}-\sin \gamma_{i a i m} g_{i j a m}^{\prime}\right)
\end{array}\right. \tag{14}
\end{gather*}

Then, substituting (13) to (12) and exchanging the order of the summation symbol, a more detailed active injection equation with Va decoupled can be obtained in (15).

\begin{equation*}
P_{i a} \approx \sum_{m \in \Omega}\binom{h_{i i a m} U_{i a} U_{i m}+\sum_{j=1, j \neq i}^{n} h_{i j a m} U_{i a}\left(U_{i m}-U_{j m} \cos \theta_{i m j m}\right)}{-\sum_{j=1, j \neq i}^{n} f_{i j a m} U_{i a} U_{j m} \sin \theta_{i m j m}} \tag{15}
\end{equation*}

Assumption 1 (Eq. (11), Phase Angle Difference Constancy) is not merely a minor simplification. It is a physics-informed model transformation that leverages the known, quasi-constant phase relationships in both the three-phase grid and the two-phase traction network. By applying this assumption before linearization, we reconstruct the problem from Eq. (2) into the more tractable form of Eq. (15). This critical step pre-decouples the phase angles, effectively bypassing the most convoluted part of a direct Taylor expansion.

\section*{C. Stage 2: Phase Assumption of the Branch and Linearization Approach}
The VAD at both ends of a transmission line is generally small and approximately zero in approximate calculations. Transformers in power systems are typically connected in a Y-Y configuration, with a small phase difference between the high and low-voltage sides. However, there exist significant differences in traction transformer branches, mainly determined by the type of traction transformer and the connection mode. For a V/x6 connected transformer, the angle of the contact line leads the corresponding power grid phase voltage by $30^{\circ}$, while for a Scott transformer, it is $60^{\circ}$, as shown in Table I. On these branches, it is not valid to assume $\cos \theta_{i j} \approx 1$ and $\sin \theta_{i j} \approx \theta_{i}-\theta_{j}$ as in the traditional linearization approach. Therefore, to maintain generality, it is assumed that the VAD at the same phase at both ends of a branch is a constant, namely Assumption 2:

\begin{equation*}
\varphi_{i m j m} \approx \theta_{i m}-\theta_{j m}, t=\theta_{i m}-\theta_{j m}-\varphi_{i m j m} \approx 0 \tag{16}
\end{equation*}

where $\varphi_{\text{imjm }}$ is a constant.\\
Based on the assumption of (16), we propose a linearization mode of sine, as shown in (17).

\begin{align*}
& \sin \left(\theta_{i m}-\theta_{j m}\right) \\
& =\sin \left(\theta_{i m}-\theta_{j m}-\varphi_{i m j m}+\varphi_{i m j m}\right)=\sin \left(t+\varphi_{i m j m}\right) \\
& =\sin t \cos \varphi_{i m j m}+\cos t \sin \varphi_{i m j m}  \tag{17}\\
& \approx t \cos \varphi_{i m j m}+\sin \varphi_{i m j m} \\
& =\left(\theta_{i m}-\theta_{j m}\right) \cos \varphi_{i m j m}-\varphi_{i m j m} \cos \varphi_{i m j m}+\sin \varphi_{i m j m}
\end{align*}

If the per unit values are used for PFC, only the $\alpha$ side of the V/x6 traction transformer is considered, we have,

\[
\left\{\begin{array}{cc}
U_{i m}=1-\Delta U_{i m} & m \neq c  \tag{18}\\
U_{i m} \approx 0 & m=c, i \in T S S
\end{array}\right.
\]

Under the premise in (18), this paper proposes the following method for decoupling Vm and Va.

\begin{align*}
& U_{i a}\left(U_{i m}-U_{j m} \cos \theta_{i m j m}\right) \quad(m \neq c) \\
& \approx U_{i a}\left(U_{i m}-U_{j m} \cos \varphi_{i m j m}\right) \\
& =\left(1-\Delta U_{i a}\right)\left(1-\Delta U_{i m}-\cos \varphi_{i m j m}\left(1-\Delta U_{j m}\right)\right)  \tag{19}\\
& \approx 1-\cos \varphi_{i m j m}-\left(2-\cos \varphi_{i m j m}\right) \Delta U_{i m}+\cos \varphi_{i m j m} \cdot \Delta U_{j m} \\
& =\left(2-\cos \varphi_{i m j m}\right) U_{i m}-\cos \varphi_{i m j m} \cdot U_{j m}+\cos \varphi_{i m j m}-1
\end{align*}

In the approximation process in (19), the second-order infinitesimal quantities $\Delta U_{i a} \cdot \Delta U_{i m}$ and $\Delta U_{i a} \cdot \Delta U_{j m}$ are considered to be approximately zero.

Based on the two approximate formulas, the hybrid phase linear decoupled active power injection equation can be derived by substituting (17) and (19) into (15), as shown in (20).

\begin{equation*}
\resizebox{0.98\columnwidth}{!}{$\displaystyle
P_{i a} \approx \sum_{m \in \Omega}\binom{h_{i a m} U_{i m}+\sum_{j=1, j \neq i}^{n} h_{i j a m}\binom{\left(2-\cos \varphi_{i m j m}\right) U_{i m}}{-\cos \varphi_{i m j m} \cdot U_{j m}+\cos \varphi_{i m j m}-1}}{-\sum_{j=1, j \neq i}^{n} f_{i j a m}\left(\left(\theta_{i m}-\theta_{j m}\right) \cos \varphi_{i m j m}-\varphi_{i m j m} \cos \varphi_{i m j m}+\sin \varphi_{i m j m}\right)}
$}\tag{20}
\end{equation*}

Further,

\begin{equation*}
\resizebox{0.98\columnwidth}{!}{$\displaystyle
\begin{aligned}
P_{i a}
&=\sum_{m \in \Omega}\left(\begin{array}{l}
\left(\sum_{j=1}^{n} h_{i j a m}\left(2-\cos \varphi_{i m j m}\right)\right) U_{i m}+\sum_{j=1, j \neq i}^{n}\left(-\cos \varphi_{i m j m} \cdot h_{i j a m}\right) U_{j m} \\
\left.-\left(\sum_{j=1, j \neq i}^{n} f_{i j a m} \cos \varphi_{i m j m}\right) \theta_{i m}+\sum_{j=1, j \neq i}^{n}\left(-f_{i j a m} \cos \varphi_{i m j m}\right) \theta_{j m}\right)+ \\
\sum_{j=1, j \neq i}^{n} h_{i j a m}\left(\cos \varphi_{i m j m}-1\right)-\sum_{j=1, j \neq i}^{n} f_{i j a m}\left(\sin \varphi_{i m j m}-\varphi_{i m j m} \cos \varphi_{i m j m}\right)
\end{array}\right)
\end{aligned}
$}\tag{21}
\end{equation*}

\begin{gather*}
=\sum_{j=1}^{n} \sum_{m \in \Omega}\left(H_{i j a m}^{\prime} U_{j m}-F_{i j a m}^{\prime} \theta_{j m}+P_{i j a m o}\right) \quad(m \neq c) \\
H_{i j a m}^{\prime}=\left\{\begin{array}{cc}
-\cos \varphi_{i m j m} \cdot h_{i j a m} & i \neq j \\
\sum_{k=1}^{n} h_{i k a m}\left(2-\cos \varphi_{i m k m}\right) & i=j
\end{array}\right.  \tag{22}\\
F_{i j a m}^{\prime}= \begin{cases}-\cos \varphi_{i m j m} f_{i j a m} & i \neq j \\
\sum_{k=1, k \neq i}^{n} \cos \varphi_{i m k m} f_{i k a m} & i=j\end{cases}  \tag{23}\\
P_{i j a m o}=h_{i j a m}\left(\cos \varphi_{i m j m}-1\right)-f_{i j a m}\left(\sin \varphi_{i m j m}-\varphi_{i m j m} \cos \varphi_{i m j m}\right) \tag{24}
\end{gather*}

On the other hand, according to the branch admittance matrix of the V/x6 traction transformer, there are

\begin{align*}
& h_{i j a m}=f_{i j a m}=0 \Rightarrow H_{i j a m}^{\prime}=F_{i j a m}^{\prime}=P_{i j a m o}=0  \tag{25}\\
& (m=c, \quad i, j \in \mathrm{~V} / \times 6 \text{ transformer })
\end{align*}

As shown in Fig. 6, since the rail nodes are either grounded or merged, the rail voltage is not involved in the calculation of (15). Therefore, it still holds true that

\begin{equation*}
P_{i a}=\sum_{j=1}^{n} \sum_{m \in \Omega}\left(H_{i j a m}^{\prime} U_{j m}-F_{i j a m}^{\prime} \theta_{j m}+P_{i j a m o}\right) \quad(m=c) \tag{26}
\end{equation*}

Combining (20) and (26), the final linear active power injection equation of phase $A$ can be obtained, as shown in (27).

\begin{equation*}
P_{i a}=\sum_{j=1}^{n} \sum_{m \in \Omega}\left(H_{i j a m}^{\prime} U_{j m}-F_{i j a m}^{\prime} \theta_{j m}+P_{i j a m o}\right) \tag{27}
\end{equation*}

Assumption 2 combined with the linearized expressions (17) and (19), further decouples voltage magnitudes. The combined use of these assumptions represents a tailored simplification scheme for our specific hybrid system.

Similarly, the linear reactive power injection equation of phase A can be obtained as follows.

\begin{gather*}
Q_{i a} \approx-\sum_{j=1}^{n} \sum_{m \in \Omega}\left(H_{i j a m}^{\prime \prime} \theta_{j m}+F_{i j a m}^{\prime \prime} U_{j m}+Q_{i j a m o}\right)  \tag{28}\\
H_{i j a m}^{\prime \prime}= \begin{cases}-\cos \varphi_{i m j m} \cdot h_{i j a m} & i \neq j \\
\sum_{k=1, k \neq i}^{n} \cos \varphi_{i m k m} h_{i k a m} & i=j\end{cases}  \tag{29}\\
F_{i j a m}^{\prime \prime}=\left\{\begin{array}{cc}
-\cos \varphi_{i m j m} f_{i j a m} & i \neq j \\
\left(\sum_{k=1}^{n} f_{i k a m}\left(2-\cos \varphi_{i m k m}\right)\right) & i=j
\end{array}\right.  \tag{30}
\end{gather*}

\begin{equation*}
\resizebox{0.98\columnwidth}{!}{$\displaystyle
Q_{i j a m o}=f_{i j a m}\left(\cos \varphi_{i m j m}-1\right)+h_{i j a m}\left(\sin \varphi_{i m j m}-\varphi_{i m j m} \cos \varphi_{i m j m}\right)
$}\tag{31}
\end{equation*}

So far, three aspects of decoupling have been achieved compared to the original node power injection equation (2): 1) coupling between voltage magnitudes; 2) coupling between voltage angles; 3) coupling between voltage magnitudes and angles. Therefore, equations (27) (28) are completely linear.

We now explicitly state: The HPLD method can be viewed as a Two-Stage Approximated Taylor Expansion: Stage 1 applies a physics-based transformation (via Assumption 1) to obtain a decoupled intermediate model (Eq. (15)); Stage 2 performs a simplified linearization on this intermediate model. This pathway is fundamentally different and more efficient than a brute-force Taylor expansion on the original model.

In conclusion, the three-phase power flow equations can be formulated with related to the PQ node, encompassing a total of six equations for three phases.

\section*{IV. Processing of PV and Slack Nodes in Three-Phase Asymmetric Power Flow Calculations}
The precise three-phase asymmetrical PFC differs from single-phase PFC when dealing with PV nodes and slack nodes. This section will give an approach to address this difference.

\section*{A. Processing of PV Nodes}
The power and voltage conditions given at the three-phase PV node can be understood as the maintenance of a specified total power $P_{e}$ and the maintenance of a specified A-phase

\begin{figure}[H]
\begin{center}
  \includegraphics[alt={},max width=\columnwidth]{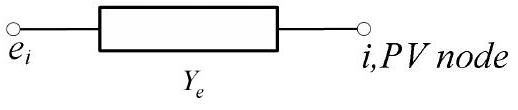}
\caption{Diagram of adding internal potential node to PV node.}
\end{center}
\end{figure}

voltage $U_{g 0}$. Therefore, these two conditions are considered insufficient for solving the three-phase state variables at the PV node. One approach that can be taken is the addition of an internal electric potential node $e_{i}$ within the generator [42], as illustrated in Fig. 7, where $Y_{e}$ represents the generator's three-phase admittance.

After adding the internal potential node $e_{i}$, the original PV node has transformed into an intermediate PQ node with zero power injection, and 6 power injection equations can be formulated. As for the internal potential node $e_{i}$, the voltage is determined by the excitation system and is three-phase balanced, that is

\begin{gather*}
U_{e_{i} a}=U_{e_{i} b}=U_{e_{i} c}  \tag{32}\\
\theta_{e_{i} b}=\theta_{e_{i} a}-2 \pi / 3, \theta_{e_{i} c}=\theta_{e_{i} a}+2 \pi / 3 \tag{33}
\end{gather*}

Therefore, for the internal potential node $e_{i}$, there are only two variables to be solved. According to the given three-phase power conditions, there is

\begin{equation*}
P_{e_{i}}=P_{e_{i} a}+P_{e_{i} b}+P_{e_{i} c} \tag{34}
\end{equation*}

By incorporating the six equations of the original PV node, a total of seven equations can be obtained. There are eight unknown variables to be solved, including the six variables of the original PV node and the two variables of the internal potential node. Additionally, based on the given voltage condition that $U_{i a}$ equals the prescribed voltage $U_{g 0}, U_{i a}$ can be eliminated from the set of unknown variables, resulting in a system with only seven unknown variables. Seven equations have been successfully established, matching the number of unknown variables, thus enabling the system to be solved.

\section*{B. Processing of Slack Node}
In the calculation of asymmetrical three-phase power flow, only the voltage amplitude and angle of phase A at the slack node can be fixed, that is

\begin{equation*}
U_{R A}=U_{R 0}, \theta_{R A}=\theta_{R 0} \tag{35}
\end{equation*}

where $U_{R A}$ and $\theta_{R A}$ represent the Vm and Va of phase A of the slack node, respectively, and $U_{R 0}$ and $\theta_{R 0}$ represent the given value. The Vm and Va of the other two phases at the slack node are the four unknown variables. A similar approach to the PV node is adopted.

In summary, a complete set of linear power flow equations for a hybrid asymmetrical ATPGS can be obtained, which represents the primary objective of this study.

\section*{V. Case Study}
The existing standard cases (such as IEEE-39, IEEE-118, etc.) mostly only include the PG topology without incorporating the TN, making it impossible to analyze the research content of this study using the above standard test cases. Therefore, the author takes the actual ATPGS in a certain region of China as a research case. The topological structure of

\begin{figure}[H]
\begin{center}
  \includegraphics[alt={},max width=\columnwidth]{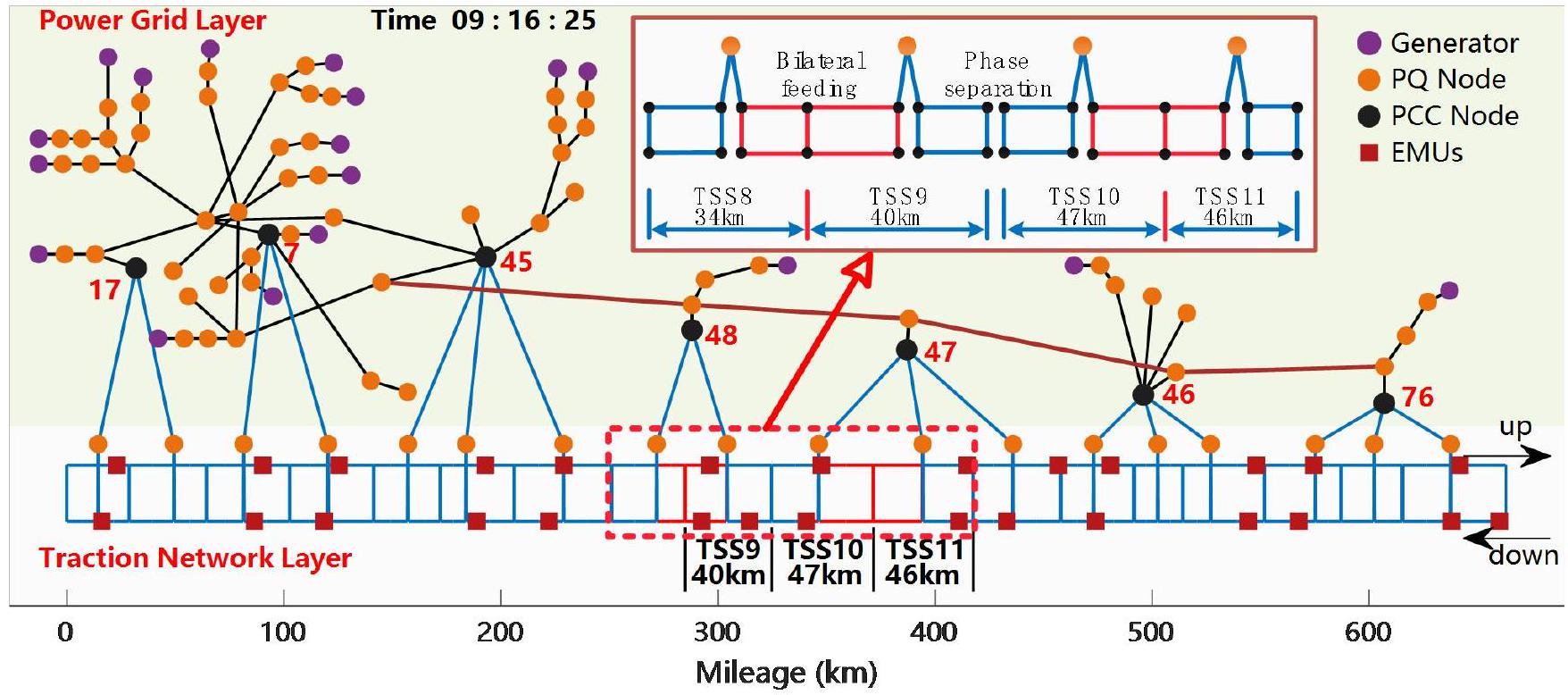}
\caption{Schematic Diagram of the ATPGS Topology Structure with EMUs.}
\end{center}
\end{figure}

\begin{figure}[H]
\begin{center}
  \includegraphics[alt={},max width=\columnwidth]{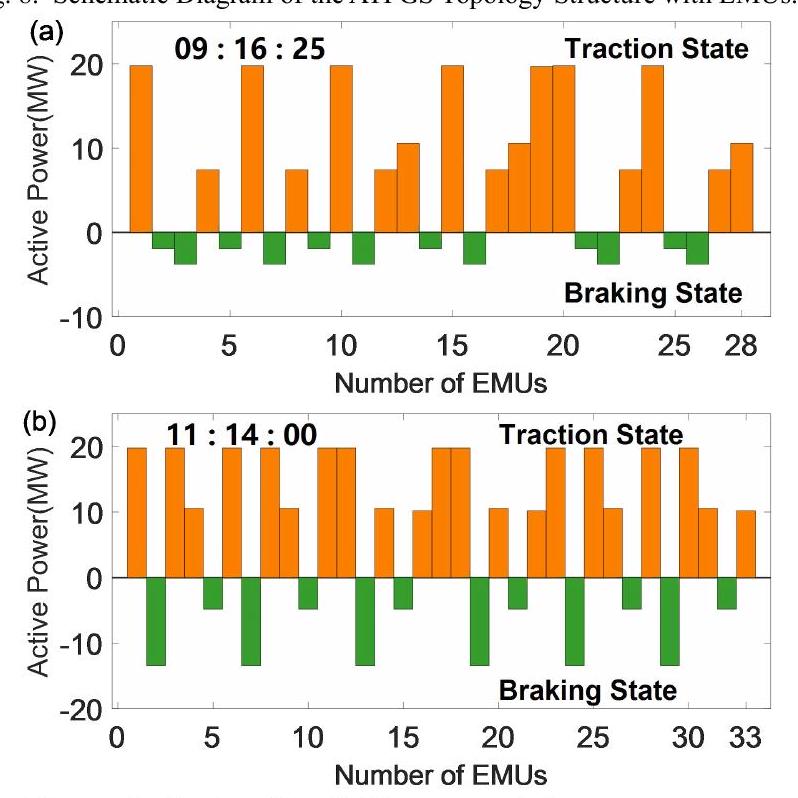}
\caption{Power distribution of the EMUs at two typical moments.}
\end{center}
\end{figure}

this ATPGS is shown in Fig. 8, which includes 18 TSSs, and a total railway length of 664 km . The upper layer represents the PG, which consists of 99 three-phase nodes, including 1 slack node, 17 PV nodes, and 81 PQ nodes, with a total load of 801 MW. There are 99 branches. The lower layer represents the TN layer, which contains a total of 18 TSSs, numbered sequentially from left to right. Each TSS has two supply arms, making a total of 36 supply arms. The supply range for TSS8 is 34 km , for TSS9 is 40 km , for TSS10 is 47 km , and for TSS11 is 46 km . In order to ensure that the EMUs do not lose power on the long slopes, a bilateral power supply method is used between TSS8 and TSS9, and also between TSS10 and TSS11, meaning the phase separation between the right supply arm of TSS8 and the left supply arm of TSS9 is eliminated, allowing the two supply arms to connect directly. Two topological structures are designed for analysis, as shown in Table S-I and S-II of the supplementary materials, including the PCC for connecting the

TSSs, along with the connection phase sequence and the lengths of the arms. Topology 1 employs V/x6 traction transformers exclusively, while Topology 2 alternates between $\mathrm{V} / \mathrm{x} 6$ and $\mathrm{V} / \mathrm{x} 0$ traction transformers, with the traction substations connected to the grid via phase rotation.

When there is no EMUs in the TSSs, the entire topology consists of 351 three-phase nodes. When the EMUs are running on the traction network, their positions continually change, resulting in ongoing modifications to the topology of the TN and the number of nodes. The operational conditions for the EMUs are set according to the timetable. At 9:16:25 of the timetable, there are a total of 28 EMUs operating in the TN, as illustrated in Fig. 8, resulting in a total of 379 three-phase nodes. EMUs are mainly in traction mode with minimal braking power, representing a typical high-load traction scenario, as shown in Fig. 9(a). At 11:14:00, there are 33 EMUs operating in the system, resulting in the total number of three-phase nodes to 384. Multiple EMUs are in downhill braking mode, representing a scenario with significant regenerative power injection, as shown in Fig. 9 (b). The active power of the EMUs at the two moments are shown in Table S-III of the supplementary materials, with a total active load of 180 MW at 09:16:25 and 222 MW at 11:14:00.

The case study is implemented in MATLAB. The desktop computer is equipped with an $\operatorname{Intel}(\mathrm{R})$ Core (TM) i5-8400 @ 2.8 GHz CPU and 24 GB of memory.

\section*{A. Voltage Unbalance Analysis at PCC}
Traction loads are asymmetric loads. In the planning of TSSs, the voltage unbalance at PCC is an important assessment indicator. Currently, design institutes generally design TSSs based solely on the short-circuit capacity at the PCC. When doing PFCs, only $1 \sim 3$ TSSs connected to the same PCC are considered. However, under the long-chain topology of mountain PG, neglecting the influence of adjacent TSSs can lead to significant errors in the calculation of voltage unbalance at the PCC. Therefore, it is necessary to consider all the TSSs

\begin{figure}[H]
\begin{center}
  \includegraphics[alt={},max width=\columnwidth]{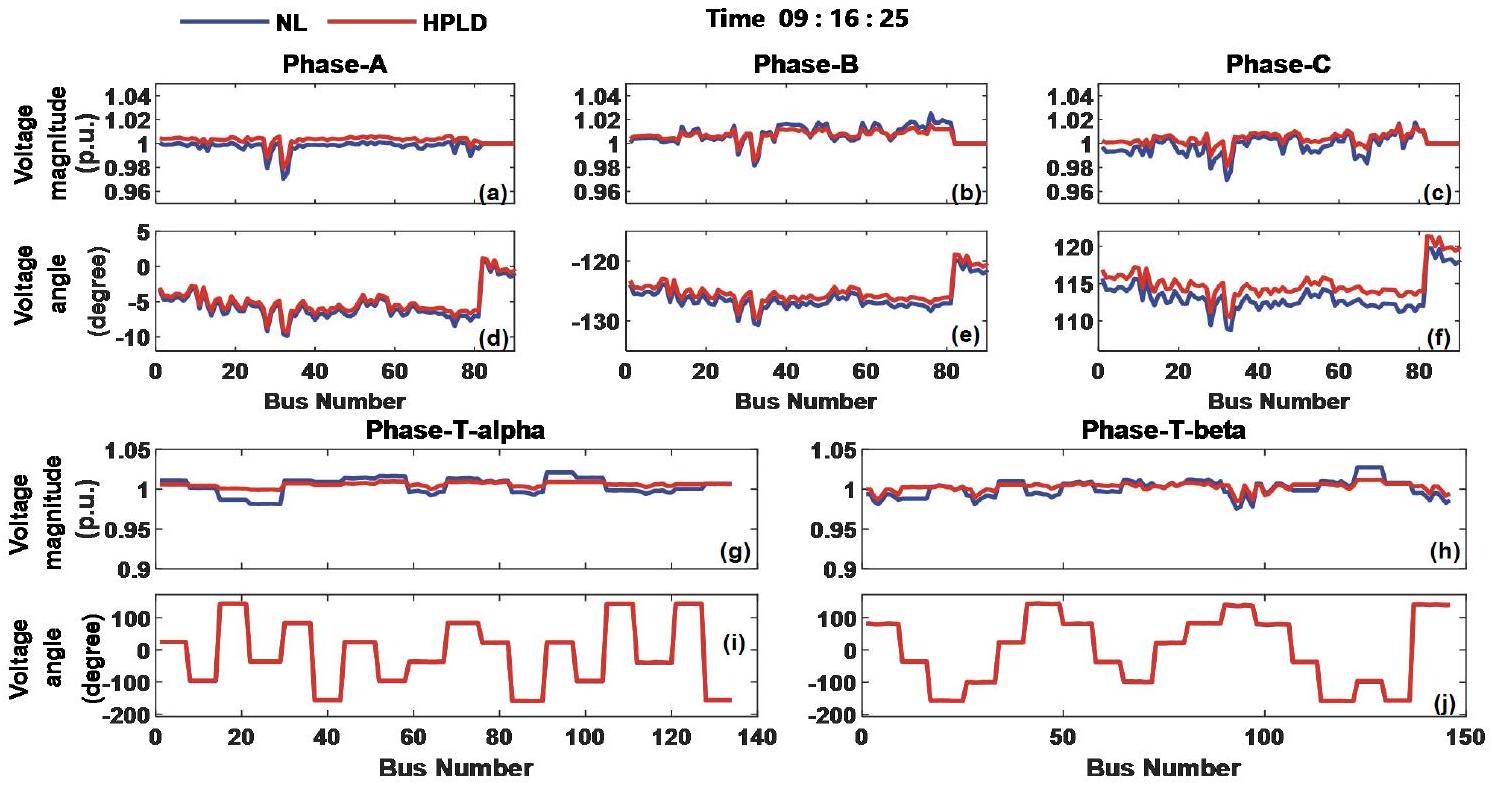}
\caption{Results of Vm and Va of ATPGS at 09:16:25.}
\end{center}
\end{figure}

\begin{table}[H]
\begin{center}
\caption{The Deviations of Voltage Unbalance at PCC Between Case 1 and Case 2}
\begin{tabular}{|l|l|l|l|l|l|l|l|}
\hline
\multicolumn{3}{|c|}{Time} & \multicolumn{5}{|c|}{09:16:25} \\
\hline
PCC Number & 17 & 7 & 45 & 48 & 47 & 46 & 76 \\
\hline
Case1(\%) & 0.81 & 0.5 & 0.63 & 0.53 & 0.44 & 0.58 & 1.07 \\
\hline
Case2(\%) & 1.24 & 1.07 & 1.02 & 0.96 & 0.72 & 0.96 & 1.07 \\
\hline
Absolute Deviation (\%) & -0.44 & -0.58 & -0.39 & -0.44 & -0.28 & -0.38 & 0 \\
\hline
Relative Deviation (\%) & -35.2 & -53.7 & -38.2 & -45.4 & -38.9 & -39.6 & 0 \\
\hline
Time & \multicolumn{7}{|c|}{11:14:00} \\
\hline
PCC Number & 17 & 7 & 45 & 48 & 47 & 46 & 76 \\
\hline
Case1(\%) & 0.92 & 0.42 & 1.05 & 0.75 & 0.99 & 0.88 & 0.9 \\
\hline
Case2(\%) & 1.2 & 0.94 & 1.18 & 1.03 & 0.7 & 0.97 & 0.63 \\
\hline
Absolute Deviation (\%) & -0.28 & -0.52 & -0.13 & -0.28 & 0.29 & -0.09 & 0.27 \\
\hline
Relative Deviation (\%) & -23.3 & -55.3 & -11.0 & -27.2 & 41.4 & -9.3 & 42.9 \\
\hline
\end{tabular}
\end{center}
\end{table}

\begin{table}[H]
\begin{center}
\caption{Traction load conditions at different time points}
\begin{tabular}{|l|l|l|l|}
\hline
Topology & Time point & Active power of EMUs (MW) & Operating scenario \\
\hline
Topology1 & 11:14:00 & 222 & Peak mixed load \\
\hline
Topology1 & 09:16:25 & 180 & Heavy traction load \\
\hline
Topology2 & 10:00:00 & 119 & Moderate traction load \\
\hline
Topology2 & 11:00:00 & 80 & Light traction load \\
\hline
Topology2 & 13:00:00 & -124 & Regenerative braking load \\
\hline
\end{tabular}
\end{center}
\end{table}

along the entire railway trunk line and the long-chain power grid for analysis. Thu, this paper analyzes the voltage unbalance at 7 PCC of the ATPGS at two heavy load times: 09:16:25 and 11:14:00 (when the load is at its maximum). Two cases are applied at each PCC.

Case 1: Only consider traction loads connected at the same PCC, equating the upper-level power grid to a single node and not taking into account other loads on the entire railway.

Case 2: Use the coupling system power flow calculation method proposed in this paper, considering all unbalanced traction loads on the entire railway.

The results of the two cases are shown in Table II, where the deviation is based on case 2. It can be seen that in long-chain power networks, when not considering other traction loads, the calculation deviation at the PCC can be significant, with both\\
positive and negative deviations possible. For example, at 11:14:00, the absolute deviation of PCC 7 reached $-0.52 \%$, and the relative deviation reached $-55.3 \%$. The relative deviation of PCC 47 and 76 reached $41.4 \%$ and $42.9 \%$, respectively. According to IEC/TR 61000-3-13, the voltage unbalance at PCC does not exceed $2 \%$. If the grid is even weaker, the results of case 1 may not meet the standard requirements. The significant deviations observed (e.g., a relative deviation of -55.3\% for PCC 7 at 11:14:00) quantitatively demonstrate that neglecting systemic coupling effects leads to severe underestimation or overestimation of voltage unbalance at PCCs. This directly justifies the necessity of the proposed integrated model for the planning and operation of ATPGS in mountainous areas-an issue that traditional methods fail to address.

\section*{B. Sensitivity Analysis Accuracy Verification of the HPLD}
The proposed HPLD and the NR method are performed for flow calculation and analysis of the ATPGS at two heavy load moments: 09:16:25 and 11:14:00 under Topology 1 and three o'clock time points: 10:00:00, 11:00:00 and 13:00:00 under Topology 2. The traction load conditions at different time points are shown in Table III. The Vm and Va belonging to the PG and TSS under the two algorithms are recorded, respectively. The errors of the Vm and Va are formulated as

\begin{align*}
& \operatorname{DUM}_{S D}^{m}=\frac{1}{N_{S D}} \sum_{k=1}^{N_{S D}}\left|U_{k m}^{H P L D}-U_{k m}^{N R}\right| \\
& \operatorname{DUA}_{S D}^{m}=\frac{1}{N_{S D}} \sum_{k=1}^{N_{S D}}\left|\theta_{k m}^{H P L D}-\theta_{k m}^{N R}\right|  \tag{36}\\
& m \in\{a, b, c\}, S D \in\{\mathrm{PG}, \mathrm{TSS}\}
\end{align*}

where $\mathrm{DUM}_{S D}^{m}$ and $\mathrm{DUA}_{S D}^{m}$ represent the average error of the Vm and Va belonging to the PG or TSS. $N_{S D}$ is the number of nodes on the PG or TSS side.

The results at 09:16:25 are shown in Fig. 10, where (a)\~{}(f) represent the Vm and Va on the PG side, (g) $\sim(\mathrm{j})$ represent the

\begin{table}[H]
\begin{center}
\caption{The average error of Vm and Va}
\begin{tabular}{|l|l|l|l|l|l|}
\hline
Time & \multicolumn{5}{|c|}{09:16:25 (Topology 1)} \\
\hline
Phase & A & B & C & T- $\alpha$ & T- $\beta$ \\
\hline
DUM(p.u.) & 0.0043 & 0.0024 & 0.0046 & 0.0068 & 0.0060 \\
\hline
DUA(deg) & 0.58 & 1.04 & 1.70 & 1.19 & 1.29 \\
\hline
Time & \multicolumn{5}{|c|}{11:14:00 (Topology 1)} \\
\hline
Phase & A & B & C & $\mathrm{T}-\alpha$ & T- $\beta$ \\
\hline
DUM(p.u.) & 0.0031 & 0.0025 & 0.0083 & 0.0075 & 0.0082 \\
\hline
DUA(deg) & 0.71 & 1.91 & 1.60 & 1.55 & 1.61 \\
\hline
Time & \multicolumn{5}{|c|}{10:00:00 (Topology 2)} \\
\hline
Phase & A & B & C & T- $\alpha$ & T- $\beta$ \\
\hline
DUM(p.u.) & 0.0024 & 0.0028 & 0.0045 & 0.0071 & 0.0051 \\
\hline
DUA(deg) & 0.09 & 0.82 & 1.25 & 0.73 & 0.80 \\
\hline
Time & \multicolumn{5}{|c|}{11:00:00 (Topology 2)} \\
\hline
Phase & A & B & C & T- $\alpha$ & T- $\beta$ \\
\hline
DUM(p.u.) & 0.0017 & 0.0032 & 0.0051 & 0.0065 & 0.007 \\
\hline
DUA(deg) & 0.39 & 0.91 & 1.01 & 0.55 & 0.54 \\
\hline
Time & \multicolumn{5}{|c|}{13:00:00 (Topology 2)} \\
\hline
Phase & A & B & C & $\mathrm{T}-\alpha$ & T- $\beta$ \\
\hline
DUM(p.u.) & 0.0018 & 0.0014 & 0.0024 & 0.0036 & 0.0035 \\
\hline
DUA(deg) & 0.2871 & 0.8747 & 0.9945 & 0.7376 & 0.7428 \\
\hline
\end{tabular}
\end{center}
\end{table}

TSS side. It should be noted that the horizontal axis in the figure does not represent node numbers but node quantities. It can be observed that, whether on the PG side or the TSS side, the proposed HPLD method maintains excellent consistency with the NR method in terms of Vm and Va calculations across all operating scenarios-covering light traction, peak traction, and regenerative braking conditions.

The average errors under different load levels and transformer configurations are summarized in Table IV. Specifically, the five simulated time points represent diverse operational states: 09:16:25 (Topology 1, 180 MW , traction-dominated), 11:14:00 (Topology 1, 222 MW , peak mixed load with multiple EMUs braking), 10:00:00 (Topology 2, 119 MW , moderate traction), 11:00:00 (Topology 2, 80 MW , light traction), and 13:00:00 (Topology 2, -124 MW, large-scale regenerative braking with net negative load). Among these, the overall errors of all indicators at 11:14:00 (peak load) are slightly higher than those at other time points, which is attributed to the more significant system nonlinearity under heavy load conditions.

For Vm calculation: The maximum average error (DUM) on the PG side occurs in phase C at 11:14:00, reaching 0.0083 p.u. ( $0.83 \%$ of the rated voltage), while the maximum average DUM on the TSS side is 0.0082 p.u. (T- $\beta$ side at 11:14:00). For lighter load scenarios (e.g., 11:00:00 and 13:00:00 in Topology 2), the average DUM is as low as 0.0017 p.u. (PG phase A) and 0.0035 p.u. (TSS T- $\beta$ side), respectively. All Vm errors are within $0.83 \%$ of the rated voltage. Such small voltage magnitude errors directly ensure the accuracy of reactive power calculation. Since reactive power transmission is linearly correlated with voltage magnitude differences between nodes, the high precision of voltage magnitude calculation confirms that the HPLD method can accurately capture reactive power flow characteristics, even with the adopted approximations.

For Va calculation: The maximum average error (DUA) on the PG side is $1.91^{\circ}$ (phase B at 11:14:00), and the maximum average DUA on the TSS side is $1.61^{\circ}$ ( $\mathrm{T}-\beta$ side at $11: 14: 00$ ). In contrast, under light load and regenerative braking conditions,\\
the average DUA is minimized to $0.09^{\circ}$ (PG phase A at 10:00:00) and $0.54^{\circ}$ (TSS T- $\beta$ side at 11:00:00), respectively. These small Va errors (all $\leq 1.91^{\circ}$ ) well guarantee the calculation accuracy of active power, even during reverse power flow caused by regenerative braking.

To summarize, compared to the nonlinear NR method, the proposed HPLD method achieves high computational accuracy within the acceptable range for engineering applications across diverse load levels ( 80 MW to 222 MW ) and operational modes (traction and regenerative braking). Meanwhile, it demonstrates robust compatibility with both pure V/x6 (Topology 1) and mixed V/x6-V/x0 (Topology 2) traction transformer configurations, confirming its strong stability and adaptability to the significant load fluctuations of electrified railway systems.

\section*{C. Validation of Key Assumptions for Linearization}
To ensure the reliability of the proposed HPLD power flow method, this section validates the two core assumptions for linearization-constant voltage angle differences between phases at the same node and across branches-using actual operating data from the ATPGS case. Sensitivity analyses are conducted covering 18 traction substations (TSSs) and two extreme operating conditions to verify the assumptions' applicability under dynamic and variable scenarios.

\begin{itemize}
  \item Traction-dominated condition (09:16:25): Most EMUs operate in traction mode with minimal braking power, as shown in Fig. 9(a).
  \item Braking-dominated condition (11:14:00): Multiple EMUs operate in downhill braking mode, as shown in Fig. 9(b).\\
The assumption of constant voltage angle differences between phases at the same node is derived from the inherent characteristics of the power system and AT traction network: three-phase grid nodes maintain a $120^{\circ}$ phase difference, while T-phase and F-phase of the AT traction network maintain a $180^{\circ}$ phase difference determined by the V/x6 traction transformer. To validate this, phase angle data of 99 three-phase grid nodes and 18 TSSs ( $\alpha$ and $\beta$ sides) at two typical moments (09:16:25 and 11:14:00) are analyzed, with key results shown in Tables V and VI.
\end{itemize}

For three-phase grid nodes: At 09:16:25, the average angle (AA) differences $\theta_{\mathrm{AB}}, \theta_{\mathrm{BC}}$, and $\theta_{\mathrm{CA}}$ are $120.6^{\circ}, 120.5^{\circ}$, and $118.9^{\circ}$, respectively, with deviations from the ideal $120^{\circ}$ no more than $1.1^{\circ}$. At 11:14:00, the corresponding average values are $120.7^{\circ}, 119.7^{\circ}$, and $119.5^{\circ}$, with maximum deviations within $1.0^{\circ}$. These results confirm that the phase angle differences of grid nodes remain stable and close to the ideal value under different operating conditions.

For AT traction network nodes: At both moments, the phase differences $\theta_{T F}$ ( T and F-phase) of all 18 TSSs are within $179.8^{\circ}-180.1^{\circ}$, with deviations from the ideal $180^{\circ}$ no more than $0.2^{\circ}$. This verifies the stability of the phase relationship between T and F-phase, which is not affected by dynamic changes in EMUs operations (e.g., traction or braking).

On the other side, the assumption of constant voltage angle differences across branches extends the traditional linearization method by considering the inherent phase shift of traction

\begin{table}[H]
\begin{center}
\caption{Calculation results of the phase angle difference at 09:16:25}
\begin{tabular}{|l|l|l|l|l|l|l|l|l|}
\hline
Time & \multicolumn{8}{|c|}{09:16:25} \\
\hline
\multirow{2}{*}{Power Grid} & variable & $\theta_{\mathrm{A}}$ & $\theta_{\mathrm{B}}$ &  & $\theta_{\text{C }}$ & $\theta_{\text{AB }}$ & $\theta_{\text{BC }}$ & $\theta_{\text{CA }}$ \\
\hline
 & AA( ${ }^{\circ}$ ) & -5.3 &  & -125.9 & 113.6 & 120.6 & 120.5 & 118.9 \\
\hline
No. TSSs & variable & $\theta_{T \alpha}$ & $\theta_{F \alpha}$ & $\theta_{T \beta}$ & $\Theta_{F \beta}$ & $\theta_{T F \alpha}$ &  & $\theta_{T F \beta}$ \\
\hline
No. 1 & AA $\left({ }^{\circ}\right)$ & 23.8 & -156.2 & 79.4 & -100.4 & 180 &  & 179.8 \\
\hline
No. 2 & AA $\left({ }^{\circ}\right)$ & -97.2 & 82.8 & -36.2 & 143.8 & 180 &  & 180 \\
\hline
No. 3 & AA( ${ }^{\circ}$ ) & 143 & -37 & -157.2 & 22.8 & 180 &  & 180 \\
\hline
No. 4 & AA( ${ }^{\circ}$ ) & -36.7 & 143.3 & -101.2 & 79 & 180 &  & 179.8 \\
\hline
No. 5 & $\mathrm{AA}\left({ }^{\circ}\right)$ & 82.3 & -97.7 & 23.3 & -156.7 & 180 &  & 180 \\
\hline
No. 6 & $\mathrm{AA}\left({ }^{\circ}\right)$ & -156.7 & 23.3 & 142.2 & -37.8 & 180 &  & 180 \\
\hline
No. 7 & $\mathrm{AA}\left({ }^{\circ}\right)$ & 23.5 & -156.6 & 79.5 & -100.5 & 180 &  & 180 \\
\hline
No. 8 & AA( ${ }^{\circ}$ ) & -97.5 & 82.5 & -37.6 & 142.4 & 180 &  & 180 \\
\hline
No. 9 & AA $\left({ }^{\circ}\right)$ & -37.6 & 142.4 & -99.4 & 80.7 & 180 &  & 179.9 \\
\hline
No. 10 & AA $\left({ }^{\circ}\right)$ & 82.7 & -97.3 & 21.3 & -158.6 & 180 &  & 180 \\
\hline
No. 11 & AA( ${ }^{\circ}$ ) & 22.2 & -157.8 & 81.6 & -98.3 & 180 &  & 179.9 \\
\hline
No. 12 & $\mathrm{AA}\left({ }^{\circ}\right)$ & -159.2 & 20.9 & 136.9 & -42.8 & 179.9 &  & 179.8 \\
\hline
No. 13 & $\mathrm{AA}\left({ }^{\circ}\right)$ & 22.7 & -157.3 & 78.1 & -101.7 & 180 &  & 179.8 \\
\hline
No. 14 & $\mathrm{AA}\left({ }^{\circ}\right)$ & -97.9 & 82.1 & -37.6 & 142.4 & 180 &  & 180 \\
\hline
No. 15 & AA( ${ }^{\circ}$ ) & 142.4 & -37.6 & -157.7 & 22.4 & 180 &  & 179.9 \\
\hline
No. 16 & AA $\left({ }^{\circ}\right)$ & -40.7 & 139.3 & -97.9 & 82.1 & 180 &  & 180 \\
\hline
No. 17 & AA( ${ }^{\circ}$ ) & 142.3 & -37.7 & -156.7 & 23.3 & 180 &  & 180 \\
\hline
No. 18 & AA $\left({ }^{\circ}\right)$ & -156.7 & 23.3 & 139.8 & -40.1 & 180 &  & 179.9 \\
\hline
\end{tabular}
\end{center}
\end{table}

\begin{table}[H]
\begin{center}
\caption{Calculation results of the phase angle difference at 11:14:00}
\begin{tabular}{|l|l|l|l|l|l|l|l|l|}
\hline
Time & \multicolumn{8}{|c|}{11:14:00} \\
\hline
\multirow{2}{*}{Power Grid} & variable & $\theta_{\mathrm{A}}$ &  & $\theta_{\mathrm{B}}$ & $\theta_{\mathrm{C}}$ & $\theta_{\text{AB }}$ & $\theta_{\mathrm{BC}}$ & $\theta_{\text{CA }}$ \\
\hline
 & AA $\left({ }^{\circ}\right)$ & -5.8 &  & -126.5 & 113.7 & 120.7 & 119.7 & 119.5 \\
\hline
No.TSSs & variable & $\theta_{T \alpha}$ & $\theta_{F \alpha}$ & $\theta_{T \beta}$ & $\theta_{F \beta}$ & $\theta_{\text{TFa }}$ &  & $\theta_{T F \beta}$ \\
\hline
No. 1 & AA( ${ }^{\circ}$ ) & 23.3 & -156.7 & 81.2 & -98.7 & 180 &  & 179.9 \\
\hline
No. 2 & AA( ${ }^{\circ}$ ) & -101.4 & 78.8 & -38.9 & 141.2 & 179.8 &  & 179.9 \\
\hline
No. 3 & AA $\left({ }^{\circ}\right)$ & 144 & -36.1 & -157.2 & 22.8 & 180.1 &  & 180 \\
\hline
No. 4 & AA $\left({ }^{\circ}\right)$ & -37 & 143 & -98.9 & 81.2 & 180 &  & 179.9 \\
\hline
No. 5 & AA( ${ }^{\circ}$ ) & 77.5 & -102.2 & 20.3 & -159.6 & 179.7 &  & 179.9 \\
\hline
No. 6 & AA $\left({ }^{\circ}\right)$ & -156.8 & 23.2 & 142.7 & -37.3 & 180.1 &  & 180 \\
\hline
No. 7 & AA( ${ }^{\circ}$ ) & 22.1 & -157.9 & 76.3 & -103.5 & 180 &  & 179.8 \\
\hline
No. 8 & AA( ${ }^{\circ}$ ) & -99.9 & 80.2 & -36.6 & 143.4 & 180 &  & 180 \\
\hline
No. 9 & AA( ${ }^{\circ}$ ) & -36.8 & 143.2 & -100.1 & 80 & 180 &  & 179.9 \\
\hline
No. 10 & AA( ${ }^{\circ}$ ) & 82 & -98 & 18.8 & -161 & 180 &  & 179.8 \\
\hline
No. 11 & AA( ${ }^{\circ}$ ) & 18.9 & -161 & 82.7 & -97.4 & 179.8 &  & 180 \\
\hline
No. 12 & AA $\left({ }^{\circ}\right)$ & -159.8 & 20.3 & 142.5 & -37.5 & 180 &  & 180 \\
\hline
No. 13 & AA( ${ }^{\circ}$ ) & 22.1 & -157.9 & 80 & -99.9 & 180 &  & 180 \\
\hline
No. 14 & AA( ${ }^{\circ}$ ) & -101.4 & 78.7 & -37.7 & 142.3 & 179.9 &  & 180 \\
\hline
No. 15 & AA( ${ }^{\circ}$ ) & 140.6 & -39.3 & -157.1 & 22.8 & 179.9 &  & 180 \\
\hline
No. 16 & AA( ${ }^{\circ}$ ) & -37.7 & 142.3 & -99 & 80.9 & 180 &  & 180 \\
\hline
No. 17 & AA( ${ }^{\circ}$ ) & 138.3 & -41.5 & -159.9 & 20.2 & 179.8 &  & 179.9 \\
\hline
No. 18 & AA( ${ }^{\circ}$ ) & -156.7 & 23.2 & 140 & -39.8 & 180.1 &  & 179.9 \\
\hline
\end{tabular}
\end{center}
\end{table}

transformers. For the ATPGS, branches are categorized into 3 types: PG transmission line branches (ABPG, BBPG, CBPG), traction transformer branches (TBTT), and traction network branches (T $\alpha \mathrm{BTS}, \mathrm{F} \alpha \mathrm{BTS}, \mathrm{T} \beta \mathrm{BTS}, \mathrm{F} \beta \mathrm{BTS}$ ). Validation results at the two typical moments are shown in Table VII.

Power grid transmission line branches: At 09:16:25, the average phase angle differences ( AAD ) of $\mathrm{ABPG}, \mathrm{BBPG}$, and CBPG are $0.92^{\circ}, 0.96^{\circ}$, and $0.98^{\circ}$, respectively. At 11:14:00,

\begin{table}[H]
\begin{center}
\caption{Calculation results of the branch phase angle difference at two moments}
\begin{tabular}{|l|l|l|l|l|l|l|l|l|}
\hline
Time & \multicolumn{4}{|c|}{09:16:25} & \multicolumn{4}{|c|}{11:14:00} \\
\hline
Branch & ABPG & BBPG & CBPG & TBTT & ABPG & BBPG & CBPG & TBTT \\
\hline
AAD $\left({ }^{\circ}\right)$ & 0.92 & 0.96 & 0.98 & 30.04 & 0.92 & 1.03 & 1.01 & 29.68 \\
\hline
No.TSSs & T $\alpha$ BTS & F $\alpha$ BTS & T $\beta$ BTS & $\mathrm{F} \beta \mathrm{BTS}$ & T $\alpha$ BTS & F $\alpha$ BTS & T $\beta$ BTS & F $\beta$ BTS \\
\hline
No. 1 & 0 & 0 & 0.5 & 0.2 & 0 & 0 & 0.3 & 0.1 \\
\hline
No. 2 & 0 & 0 & 0 & 0 & 0.5 & 0.3 & 0.3 & 0.2 \\
\hline
No. 3 & 0 & 0 & 0.2 & 0.1 & 0.1 & 0.1 & 0 & 0 \\
\hline
No. 4 & 0 & 0 & 0.4 & 0.2 & 0 & 0 & 0.7 & 0.3 \\
\hline
No. 5 & 0 & 0 & 0 & 0 & 0.7 & 0.4 & 0.3 & 0.1 \\
\hline
No. 6 & 0 & 0 & 0.2 & 0.1 & 0.1 & 0.1 & 0 & 0 \\
\hline
No. 7 & 0.1 & 0 & 0.1 & 0.1 & 0 & 0 & 0.8 & 0.4 \\
\hline
No. 8 & 0 & 0 & 0 & 0 & 0.1 & 0 & 0.1 & 0 \\
\hline
No. 9 & 0.2 & 0.1 & 0.2 & 0.2 & 0.1 & 0.1 & 0.3 & 0.2 \\
\hline
No. 10 & 0 & 0 & 0.3 & 0.2 & 0 & 0 & 0.8 & 0.3 \\
\hline
No. 11 & 0.2 & 0.2 & 0.2 & 0.1 & 0.4 & 0.3 & 0 & 0 \\
\hline
No. 12 & 0.1 & 0 & 0.8 & 0.6 & 0.1 & 0 & 0 & 0 \\
\hline
No. 13 & 0 & 0 & 0.5 & 0.2 & 0 & 0 & 0.6 & 0.2 \\
\hline
No. 14 & 0 & 0 & 0 & 0 & 0.2 & 0.1 & 0 & 0 \\
\hline
No. 15 & 0 & 0 & 0.2 & 0.1 & 0.2 & 0.1 & 0.1 & 0 \\
\hline
No. 16 & 0.1 & 0.1 & 0 & 0 & 0 & 0 & 0.3 & 0.1 \\
\hline
No. 17 & 0 & 0 & 0 & 0 & 0.5 & 0.3 & 0.3 & 0.2 \\
\hline
No. 18 & 0 & 0 & 0.4 & 0.3 & 0.2 & 0.1 & 0.5 & 0.4 \\
\hline
\end{tabular}
\end{center}
\end{table}

the corresponding values are $0.92^{\circ}, 1.03^{\circ}$, and $1.01^{\circ}$, all close to $0^{\circ}$, consistent with the theoretical characteristic of small phase angle differences in transmission lines.

Traction transformer branches: The AAD of TBTT branches are $30.04^{\circ}(09: 16: 25)$ and $29.68^{\circ}(11: 14: 00)$, nearly identical to the $30^{\circ}$ phase shift of V/x6 traction transformers, confirming the stability of the phase angle difference for this branch type.

Traction network branches: For all 18 TSSs, the AAD of $\mathrm{T} \alpha \mathrm{BTS}, \mathrm{F} \alpha \mathrm{BTS}, \mathrm{T} \beta \mathrm{BTS}$, and $\mathrm{F} \beta \mathrm{BTS}$ branches are within $0^{\circ}-0.8^{\circ}$ at both moments, indicating negligible phase angle differences and validating the assumption.

The analysis involves 18 TSSs to cover spatial variations in the ATPGS. Results show that even with significant changes in EMUs operating modes (traction vs. braking), the phase angle differences at nodes and across branches remain within the stable range specified in the assumptions. The maximum deviation from ideal values does not exceed $1.1^{\circ}$ for node phase differences and $0.8^{\circ}$ for branch phase differences (TBTT branches), confirming the assumptions' applicability under dynamic operating environments.

\section*{D. Branch VAD Analysis}
Fig. 11 shows the VAD of 99 branches on the PG side ( $\mathrm{ABPG} / \mathrm{BBPG} / \mathrm{CBPG}$ ) and 36 traction transformer T -phase branches (TBTT) calculated by the proposed HPLD algorithm at 09:16:25. This moment corresponds to Topology 1 (all V/x6 transformers) with 28 EMUs operating in traction mode (total active load $=180 \mathrm{MW}$, minimal braking power), as detailed in Section V.A. To validate the accuracy of the HPLD method in capturing branch VAD, all results are compared against the Newton-Raphson (NR) method (nonlinear baseline). Tables VIII-X summarize the average and maximum values of VAD and their errors relative to the NR method.

For PG side branches: The average VAD (AAD) calculated by HPLD is $0.92^{\circ}-0.98^{\circ}$ (Table VIII), with a maximum VAD (MAD) of $6.21^{\circ}-6.67^{\circ}$. Compared to the NR method, the

\begin{figure}[H]
\begin{center}
  \includegraphics[alt={},max width=\columnwidth]{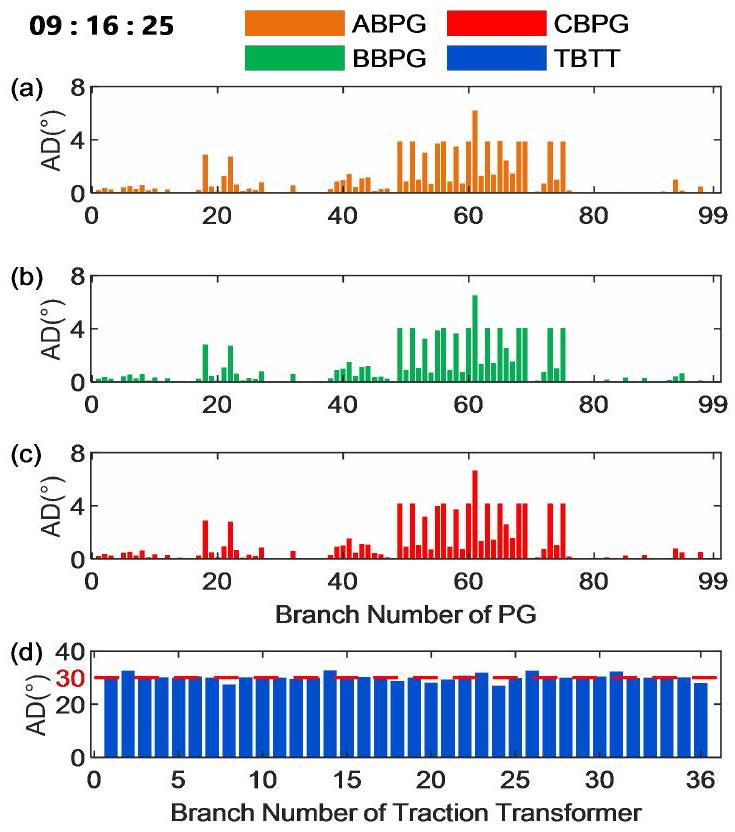}
\caption{The VAD of the branches of ATPGS.}
\end{center}
\end{figure}

\begin{table}[H]
\begin{center}
\caption{The Average and Maximum of VAD of Branches}
\begin{tabular}{ccccccccc}
\hline
Time & \multicolumn{3}{c}{$09: 16: 25$} & \multicolumn{4}{c}{$11: 14: 00$} &  \\
\hline
Branch & ABPG & BBPG & CBPG & TBTT & ABPG & BBPG & CBPG & TBTT \\
AAD $\left({ }^{\circ}\right)$ & 0.92 & 0.96 & 0.98 & 30.04 & 0.92 & 1.03 & 1.01 & 29.68 \\
MAD $\left({ }^{\circ}\right)$ & 6.21 & 6.50 & 6.67 & 32.78 & 6.19 & 7.09 & 6.92 & 33.6 \\
\hline
\end{tabular}
\end{center}
\end{table}

\section*{TABLE IX}
Va of the high and low sides of the traction transformer

\begin{center}
\begin{tabular}{|l|l|l|l|l|l|l|}
\hline
Time & \multicolumn{6}{|c|}{09:16:25} \\
\hline
Phase sequence & \multicolumn{2}{|c|}{ACB} & \multicolumn{2}{|c|}{CBA} & \multicolumn{2}{|c|}{BAC} \\
\hline
TSS number & \multicolumn{2}{|c|}{TSS1} & \multicolumn{2}{|c|}{TSS2} & \multicolumn{2}{|c|}{TSS3} \\
\hline
\multirow{2}{*}{Va of high side ( ${ }^{\circ}$ )} & \multirow{2}{*}{A -5.23} & \multirow{2}{*}{B 114.48} & \multirow{2}{*}{\begin{tabular}{l}
C \\
-125.58 \\
\end{tabular}} & \multirow{2}{*}{A -5.22} & \multirow{2}{*}{B 113.96} & C \\
\hline
 &  &  &  &  &  & -126.27 \\
\hline
\multirow{2}{*}{Va of low side ( ${ }^{\circ}$ )} & T-alfa & T-beta & T-alfa & T-beta & T-alfa & T-beta \\
\hline
 & 24.51 & 81.91 & -95.47 & -35.3 & 143.98 & -156.63 \\
\hline
TBTT $\left({ }^{\circ}\right)$ & 29.74 & 32.57 & 30.1 & 30.07 & 30.02 & 30.36 \\
\hline
\end{tabular}
\end{center}

The average and maximum value of the errors of branch VAD

\begin{table}[H]
\begin{center}
\caption{}
\begin{tabular}{ccccccccc}
\hline
Time & \multicolumn{4}{c}{$09: 16: 24$} & \multicolumn{4}{c}{$11: 14: 00$} \\
\hline
Branch & ABPG & BBPG & CBPG & TBTT & ABPG & BBPG & CBPG & TBTT \\
AAE $\left({ }^{\circ}\right)$ & 0.03 & 0.03 & 0.05 & 0.4 & 0.03 & 0.04 & 0.05 & 0.46 \\
MAE $\left({ }^{\circ}\right)$ & 0.38 & 0.68 & 1.14 & 0.73 & 0.39 & 1.32 & 1.01 & 0.83 \\
\hline
\end{tabular}
\end{center}
\end{table}

average VAD error (AAE) is $\leq 0.05^{\circ}$, and the maximum error (MAE) is $\leq 1.14^{\circ}$ (Table X), confirming the HPLD method's ability to accurately capture small phase angle differences in transmission lines.

For traction transformer branches (TBTT): The average VAD calculated by HPLD is $30.04^{\circ}$, which is nearly identical to the inherent $30^{\circ}$ phase shift of V/x6 transformers. The AAE relative to the NR method is $0.4^{\circ}$, and the MAE is $0.73^{\circ}$ (Table X ), demonstrating the method's unique advantage in handling large phase shifts-an ability that existing linear models [28]-[32] lack (as discussed in Section I.A).

According to the literature review, existing three-phase linear power flow models cannot effectively handle branches with such large phase angle differences (e.g., $30^{\circ}$ for V/x6 transformers), often simplifying $\cos \theta_{i j} \approx 1$ and leading to

\begin{table}[H]
\begin{center}
\caption{Execution Time of NR and HPLD method}
\begin{tabular}{ccc}
\hline
Time per instant & NR & HPLD \\
\hline
Maximum (s) & 32.4656 & 17.6281 \\
Minimum (s) & 13.6834 & 7.3836 \\
Mean (s) & 16.8738 & 9.4809 \\
\hline
\end{tabular}
\end{center}
\end{table}

\begin{table}[H]
\begin{center}
\caption{The Performance Comparison of Different Linearization Methods}
\begin{tabular}{|l|l|l|l|l|}
\hline
Metric & HPLD & Classical Taylor [25] & DataDriven [24] & Dimension -Rising [19] \\
\hline
Core objective & Compact linear model & Compact linear model & Modelfree & Highaccuracy linear model \\
\hline
Data dependence & No & No & Yes & No \\
\hline
Computational burden & Low & Moderate & Moderate & Low \\
\hline
Model accuracy & Moderate & Moderate & High & High \\
\hline
Asymmetry handing & Strong & Moderate & Strong & Moderate \\
\hline
\end{tabular}
\end{center}
\end{table}

significant VAD errors. In contrast, the HPLD method's VAD errors are within $1.14 \%$, fully meeting engineering calculation requirements. This validates that the proposed method fills the gap of existing linear models in handling transformer connection angles for the ATPGS.

\section*{E. Comparison of Execution Time}
To test the computational efficiency of the proposed HPLD method, the busy time from $9: 16: 25$ to $9: 46: 25$ is chosen as the experimental period. The power flow is calculated every 2 seconds. To eliminate random errors, each time profile underwent 10 times calculations, and the maximum, minimum, and average times are counted for each time profile. Table XI summarizes the statistical values of the execution time for the two methods. It shows that the execution time of the HPLD is much smaller than the NR method. For example, the average execution time for the HPLD method is 9.4809 seconds, while the computation time for the NR method is 16.8738 seconds, resulting in an efficiency improvement of approximately $44 \%$.

\section*{F. Comparative Analysis and Methodological Positioning}
The performance comparison of different linearization methods is presented in Table XII, encompassing four representative approaches: the proposed Hybrid Phase Linear Decoupled (HPLD) method, Classical Taylor Expansion [25], Data-Driven Linearization [24], and Dimension-Rising Linearization [19]. The comparison covers key evaluation dimensions, including Core Objective, Data Dependency, Computational Burden, Model Accuracy, and Asymmetry Handling. As illustrated in the table, the HPLD method is a model-driven, physics-informed simplification rooted in the Taylor expansion framework, tailored specifically for the unique characteristics of AT traction network-power grid coupled systems (ATPGS). Its standout advantages for ATPGS planning are threefold: 1) It requires no extensive training data (in contrast to data-driven methods), a critical feature for planning scenarios where comprehensive operational data is often unavailable; 2) It explicitly incorporates transformer\\
connection angles (e.g., $30^{\circ}$ for $\mathrm{V} / \mathrm{x} 6,60^{\circ}$ for Scott transformers)—a key attribute ignored by most pure power grid linear flow models; 3) It maintains high computational transparency and efficiency, enabling fast batch calculations of multiple planning scenarios.

Notably, the HPLD method retains full backward compatibility with standard power systems (without special traction transformers): when applied to conventional power grids with Y-Y transformers (negligible phase shifts), it automatically simplifies to align with classical linearization logic, achieving comparable accuracy to mainstream methods. This dual adaptability-excelling in ATPGS while remaining compatible with standard power systems-further underscores its practical engineering significance.

\section*{VI. Conclusion}
This paper addresses the planning challenges of the TSSs in remote mountainous areas characterized by a long-chain weak power grid. An integrated PFC model for the ATPGS, with a particular emphasis on the unique port load characteristics of EMUs is proposed. The model thoroughly accounts for the unbalanced characteristics inherent in the ATPGS. Additionally, the HPLD method is proposed to tackle the coupling characteristics of the ATPGS. This linear method incorporates the connection angles of the $\mathrm{V} / \mathrm{x} 6$ traction transformers, allowing for the linearization of power flow equations within the hybrid phase transmission system and thereby filling the gap in the existing literature. The linear model exhibits high accuracy in Vm while maintaining acceptable Va errors for engineering applications. Consequently, this model facilitates the conversion of nonlinear, non-convex planning models into linear programming problems for the joint planning of ATPGS, offering a robust analytical tool for planning in remote mountainous regions. It can also be extended to direct power supply systems and transmission-distribution network coupled systems. Its main limitations lie in the incomplete coverage of transformer types (Scott transformers not fully derived) and the idealized grounding assumption. The future research directions outlined above will address these constraints, further expanding the model's applicability and practical value.

\section*{References}
[1] Q. Zhang, Y. Zhang, K. Huang, et al., "Modeling of Regenerative Braking Energy for Electric Multiple Units Passing Long Downhill Section," IEEE Trans. Transp. Electrific., vol. 8, no. 3, pp. 3742-3758, 2022.\\[0pt]
[2] J. Lai, M. Chen, X. Dai, et al, "Power Flow Optimization and Control Strategy for Energy Router in Dual Mode Traction Power Supply System," IEEE Trans. Intell. Transp. Syst., vol. 24, no. 11, pp. 12284-12293, 2023.\\[0pt]
[3] Y. Liu, M. Chen, Z. Cheng, et al, "Robust Energy Management of High-Speed Railway Co-Phase Traction Substation With Uncertain PV Generation and Traction Load," IEEE Trans. Intell. Transp. Syst., vol. 23, no. 6, pp. 5079-5091, 2022.\\[0pt]
[4] B. Lu, Y. Song, Z. G. Liu, X. F. Wang, Q. Zhang, and Y. M. Lu, "Evolution Analysis of Three-Dimensional Wheel Polygonal Wear for Electric Locomotives Considering the Effect of Interharmonics," IEEE Trans. Veh. Technol., vol. 74, no. 2, pp. 2443-2457, Feb 2025.\\[0pt]
[5] K. Ponnambalam, V. H. Quintana, and A. Vannelli, "A fast algorithm for power-system optimization problems using an interior point method," IEEE Trans. Power Syst., vol. 7, no. 2, pp. 892-899, May 1992.\\[0pt]
[6] J. A. Momoh, M. E. El-Hawary, and R. Adapa, "A review of selected optimal power flow literature to 1993 part II: Newton, linear programming and interior point methods," IEEE Trans. Power Syst., vol. 14, no. 1, pp. 105-111, Feb 1999.\\[0pt]
[7] S. Shin, M. Anitescu, and F. Pacaud, "Accelerating optimal power flow with GPUs: SIMD abstraction of nonlinear programs and condensed-space interior-point methods," Electr. Power Syst. Res., vol. 236, Nov 2024, Art. no. 110651.\\[0pt]
[8] K. J. Tang, S. F. Dong, J. Shen, C. Z. Zhu, and Y. H. Song, "A Robust and Efficient Two-Stage Algorithm for Power Flow Calculation of Large-Scale Systems," IEEE Trans. Power Syst., vol. 34, no. 6, pp. 5012-5022, Nov 2019.\\[0pt]
[9] A. A. Eajal, M. A. Abdelwahed, E. F. El-Saadany, and K. Ponnambalam, "A Unified Approach to the Power Flow Analysis of AC/DC Hybrid Microgrids," IEEE Trans. Sustainable Energy, vol. 7, no. 3, pp. 1145-1158, Jul 2016.\\[0pt]
[10] M. Wang and S. S. Liu, "A trust region interior point algorithm for optimal power flow problems," Int. J. Electr. Power Energy Syst., vol. 27, no. 4, pp. 293-300, May 2005.\\[0pt]
[11] C. J. Goodman and T. Kulworawanichpong, "Sequential linear power flow solution for AC electric railway power supply systems," WIT Trans. Built Environ., vol. 61, 2002.\\[0pt]
[12] H. Hu, Z. He, X. Li, et al., "Power-Quality Impact Assessment for High-Speed Railway Associated with High-Speed Trains Using Train Timetable-Part I: Methodology and Modeling," IEEE Trans. Power Delivery, vol. 31, no. 2, pp. 693-703, 2016.\\[0pt]
[13] H. Hu, Z. He, J. Wang, et al., "Power Flow Calculation of High-speed Railway Traction Network Based on Train-network Coupling Systems," Proceedings of the CSEE, vol. 32, no. 19, pp. 101-108, 2012.\\[0pt]
[14] W. Zhang, S. Yang, J. Zhang, et al., "A two-phase power flow algorithm of traction power supply system based on traction substation-network decoupling," Int. J. Electr. Power Energy Syst., vol. 152, p. 109252, 2023.\\[0pt]
[15] B. Mohamed, P. Arboleya, I. El-Sayed, et al., "High-speed $2 \times 25 \mathrm{kV}$ traction system model and solver for extensive network simulations," IEEE Trans. Power Syst., vol. 34, no. 5, pp. 3837-3847, 2019.\\[0pt]
[16] K. Mongkoldee and T. Kulworawanichpong, "Current-based Newton-Raphson power flow calculation for AT-fed railway power supply systems," Int. J. Electr. Power Energy Syst., vol. 98, pp. 11-22, 2018.\\[0pt]
[17] H. Wang, W. Liu, Q. Li, et al., "Dynamic Power Flow Calculation of AC Electrified Railway Based on Source-grid-load Unified Chain Circuit," Proceedings of the CSEE, vol. 42, no. 11, pp. 3936-3952, 2022.\\[0pt]
[18] B. Stott, J. Jardim, and O. Alsac, "DC Power Flow Revisited," IEEE Trans. Power Syst., vol. 24, no. 3, pp. 1290-1300, 2009.\\[0pt]
[19] S. M. Fatemi, S. Abedi, G. B. Gharehpetian, et al., "Introducing a Novel DC Power Flow Method with Reactive Power Considerations," IEEE Trans. Power Syst., vol. 30, no. 6, pp. 3012-3023, 2015.\\[0pt]
[20] R. A. Jabr, "High-Order Approximate Power Flow Solutions and Circular Arithmetic Applications," IEEE Trans. Power Syst., vol. 34, no. 6, pp. 5053-5062, 2019.\\[0pt]
[21] J. Yang, N. Zhang, C. Kang, et al., "A state-independent linear power flow model with accurate estimation of voltage magnitude," IEEE Trans. Power Syst., vol. 32, no. 5, pp. 3607-3617, 2016.\\[0pt]
[22] Z. Yang, K. Xie, J. Yu, et al., "A General Formulation of Linear Power Flow Models: Basic Theory and Error Analysis," IEEE Trans. Power Syst., vol. 34, no. 2, pp. 1315-1324, 2019.\\[0pt]
[23] S. V. Dhople, S. S. Guggilam, and Y. C. Chen, "Linear approximations to AC power flow in rectangular coordinates," in 2015 53rd Annual Allerton Conference on Communication, Control, and Computing (Allerton), 2015, pp. 211-217.\\[0pt]
[24] Y. Liu, N. Zhang, Y. Wang, et al., "Data-Driven Power Flow Linearization: A Regression Approach," IEEE Trans. Smart Grid, vol. 10, no. 3, pp. 2569-2580, 2019.\\[0pt]
[25] Z. Li, J. Yu, and Q. H. Wu, "Approximate linear power flow using logarithmic transform of voltage magnitudes with reactive power and transmission loss consideration," IEEE Trans. Power Syst., vol. 33, no. 4, pp. 4593-4603, 2017.\\[0pt]
[26] E. Schweitzer, S. Saha, A. Scaglione, et al., "Lossy DistFlow Formulation for Single and Multiphase Radial Feeders," IEEE Trans. Power Syst., vol. 35, no. 3, pp. 1758-1768, 2020.\\[0pt]
[27] J. Huang, B. Cui, X. Zhou, et al., "A Generalized LinDistFlow Model for Power Flow Analysis," in 2021 60th IEEE Conference on Decision and Control (CDC), 2021, pp. 3493-3500.\\[0pt]
[28] Y. Wang, N. Zhang, H. Li, et al., "Linear three-phase power flow for unbalanced active distribution networks with PV nodes," CSEE J. Power Energy Syst, vol. 3, no. 3, pp. 321-324, 2017.\\[0pt]
[29] R. Hu, Q. Li, and F. Qiu, "Ensemble Learning Based Convex Approximation of Three-Phase Power Flow," IEEE Trans. Power Syst., vol. 36, no. 5, pp. 4042-4051, 2021.\\[0pt]
[30] I. L. Carreño, A. Scaglione, S. S. Saha, et al., "Log(v) 3LPF: A Linear Power Flow Formulation for Unbalanced Three-Phase Distribution Systems," IEEE Trans. Power Syst., vol. 38, no. 1, pp. 100-113, 2023.\\[0pt]
[31] H. Ahmadi, J. R. Martı', and A. v. Meier, "A Linear Power Flow Formulation for Three-Phase Distribution Systems," IEEE Trans. Power Syst., vol. 31, no. 6, pp. 5012-5021, 2016.\\[0pt]
[32] A. Bernstein, C. Wang, et al., "Load Flow in Multiphase Distribution Networks: Existence, Uniqueness, Non-Singularity and Linear Models," IEEE Trans. Power Syst., vol. 33, no. 6, pp. 5832-5843, 2018.\\[0pt]
[33] L. Gao, Y. H. Xu, X. N. Xiao, Y. Y. Liu, and P. S. Jiang, "Analysis of Adverse Effects on the Public Power Grid Brought by Traction Power-supply System," in IEEE Electrical Power and Energy Conference, Vancouver, CANADA, 2008, pp. 546-552, 2008.\\[0pt]
[34] Y. L. Che, X. J. Chen, H. J. Lin, Y. Gao, X. Lyu, and Y. X. Chen, "Temporal Impact of High-Speed Railway Traction Load on Grid Voltage Considering the Uncertainty," IEEE Trans. Transp. Electrif., vol. 10, no. 4, pp. 9381-9395, Dec 2024.\\[0pt]
[35] R. S. Salles, "Modeling and assessment of waveform distortion interaction at the railway grid side in low-frequency electrification systems," Int. J. Electr. Power Energy Syst., vol. 171, Oct 2025, Art. no. 111039.\\[0pt]
[36] L. X. Sun, X. Zhang, M. Liu, and H. Y. Peng, "Modeling and Influence Research of Traction Power Supply System Based on ADPSS," in International Conference on Power System Technology, Chengdu, CHINA, 2014, 2014.\\[0pt]
[37] W. L. Lyu, L. C. Cai, and D. Q. Huang, "Compressed Newton-Raphson Method for Power Flow Analysis in DC Traction Network," IEEE Trans. Power Syst., vol. 38, no. 2, pp. 1783-1786, Mar 2023.\\[0pt]
[38] M. W. Chen, Y. T. Chen, Y. Y. Chen, X. F. Dai, and L. Liu, "Unified Power Quality Management for Traction Substation Groups Connected to Weak Power Grids," IEEE Trans. Power Delivery, vol. 37, no. 5, pp. 4178-4189, Oct 2022.\\[0pt]
[39] B. An, Y. Li, J. M. Guerrero, W. J. Lee, L. Luo, and Z. Zhang, "Renewable Energy Integration in Intelligent Railway of China: Configurations, Applications and Issues," IEEE Intell. Transp. Syst. Mag., vol. 13, no. 3, pp. 13-33, 2021.\\[0pt]
[40] Z. H. He, C. Wan, H. Liu, and P. Ju, "Evaluation of Grid Support Capacity From Electrified Railways," IEEE Trans. Power Syst., vol. 39, no. 1, pp. 2310-2326, Jan 2024.\\[0pt]
[41] Z. He, H. Hu, Y. Zhang, et al., "Harmonic Resonance Assessment to Traction Power-Supply System Considering Train Model in China High-Speed Railway," IEEE Trans. Power Delivery, vol. 29, no. 4, pp. 1735-1743, 2014.\\[0pt]
[42] T. Wang, S. Wang, S. Ma, J. Guo, and X. Zhou, "An Extended Continuation Power Flow Method for Static Voltage Stability Assessment of Renewable Power Generation-Penetrated Power Systems," IEEE Trans. Circuits Syst. II Express Briefs, vol. 71, no. 2, pp. 892-896, 2024.

\end{document}